\documentclass[graybox,natbib,nosecnum]{svmult}
\bibpunct{(}{)}{;}{a}{}{,} 

\pdfoutput=1   

\usepackage{mathptmx}       
\usepackage{helvet}         
\usepackage{courier}        
\usepackage{type1cm}        

\usepackage{makeidx}         
\usepackage{graphicx}        
\usepackage{multicol}        
\usepackage[bottom]{footmisc}
\usepackage[normalem]{ulem}	
\usepackage{hyperref}  

\usepackage{amsmath}
\usepackage{amssymb}

\usepackage{soul}   

\def\mearth{{\,M_\oplus}}
\def\rearth{{\,R_\oplus}}

\makeindex             

\begin{document}

\title*{Formation of Terrestrial Planets}
\author{Andr\'e  Izidoro  and Sean N. Raymond}
\institute{Andr\'e Izidoro \at UNESP, Univ. Estadual Paulista - Grupo de Din\^amica Orbital Planetologia, Guaratinguet\'a , CEP
12.516-410, São Paulo, Brazil \email{izidoro.costa@gmail.com}
\and Sean N. Raymond \at Laboratoire d'Astrophysique de Bordeaux, Univ. Bordeaux, CNRS, B18N, allée Geoffroy Saint-Hilaire, 33615 Pessac, France\email{rayray.sean@gmail.com}}
%
%
\maketitle

\abstract{ The past decade has seen major progress in our understanding of terrestrial planet formation. Yet key questions remain. In this review we first address the growth of 100 km-scale planetesimals as a consequence of dust coagulation and concentration, with current models favoring the streaming instability. Planetesimals grow into Mars-sized (or larger) planetary embryos by a combination of pebble- and planetesimal accretion. Models for the final assembly of the inner Solar System must match constraints related to the terrestrial planets and asteroids including their orbital and compositional distributions and inferred growth timescales. Two current models -- the Grand-Tack and low-mass (or empty) primordial asteroid belt scenarios -- can each match the empirical constraints but both have key uncertainties that require further study. We present formation models for close-in super-Earths -- the closest current analogs to our own terrestrial planets despite their very different formation histories -- and for terrestrial exoplanets in gas giant systems. We explain why super-Earth systems cannot form in-situ but rather may be the result of inward gas-driven migration followed by the disruption of compact resonant chains. The Solar System is unlikely to have harbored an early system of super-Earths; rather, Jupiter's early formation may have blocked the ice giants' inward migration. Finally, we present a chain of events that may explain why our Solar System looks different than more than 99\% of exoplanet systems.
}

\section{Introduction}



Over the last three decades significant progress has been made on our understanding of how terrestrial planets form. On the theoretical side, this was in large part catalyzed by interdisciplinary scientific efforts and technological advances (e.g., faster computers, better software).  Technology has also been crucial in driving observational advances. Although this field by necessity started within the context of our Solar System, new observational and theoretical studies have provided a push toward a more general, broadly-applicable framework.
 
Radioactive dating and the determination of chemical compositions of different Solar System bodies have led to major advances. Constraints on the ages and compositions of different planets and small bodies directly connect with models of their origins and interiors.  Improvements in computational capabilities -- both in hardware to software -- have enabled more sophisticated and realistic numerical simulations that model a range of chemical and physical processes across all stages of planet formation. Modern planet formation theories are at the interface of empirical studies, geochemical and cosmochemical analyses, and dynamical simulations. Yet the formation of the terrestrial planets remains a subject of intense debate. At least two different models can explain the origin of our inner Solar System. Each model is based on different assumptions, left to be disentangled by future research.

Low-mass, potentially terrestrial exoplanets are a hot topic in astronomy. The discovery of such planets has been a major success of planet-finding missions such as NASA's Kepler mission~\citep{boruckietal10}. The search for exo-terrestrial planets is especially exciting because they are potential candidates for hosting life as we know it. To date more than 3000 exoplanets have been confirmed but observations have found that most planetary systems have dynamical architectures strikingly different from our own.  Gas giant exoplanets have been observed on orbits very different than those of Jupiter and Saturn, including very close-in hot Jupiters and on very eccentric orbits~\citep[e.g.][]{butleretal06,udryetal07}. Compact systems of hot super-Earths -- with sizes between 1 and 4 Earth radii; or masses between 1 and 20 Earth mass -- are systematically found orbiting their stars at distances much shorter than that of Mercury to our sun~\citep{howardetal10,mayoretal11}. This class of planet seems to be present around the majority of   main sequence stars~\citep[e.g.][]{howardetal12,fressinetal13,petiguraetal13} but no such planet exists in our inner Solar System. Given that most exoplanetary systems look dramatically different than our own, should we expect those planets to have formed in the same ways as our own terrestrial planets, or by completely different processes?

In this article we review terrestrial planet formation in the Solar System and around other stars. We discuss the dynamical processes that shaped the inner Solar System and our understanding of different formation pathways that can explain the orbital architectures of exoplanet systems.  We present a path toward understanding how our Solar System fits in the larger context of exoplanets, and how exoplanets themselves can be used to improve our understanding of our own Solar System.

 Any successful model of Solar System formation must explain why Mars is so much smaller than Earth and why the asteroid belt is so low in mass yet dynamically excited. There are currently two models that can explain the inner Solar System. Likewise, any successful model for the formation of super-Earths must match the observed distributions (e.g., the period ratio distribution). We will discuss the relationships between super-Earths and true `terrestrial planets', which are likely far more different than one may expect based solely on their observed sizes/masses.  We will also discuss why the Solar System is particularly outstanding in its lack hot super-Earths inside Mercury's orbit.

This review is organized as follows. We first briefly discuss the early stages of  planet formation, from dust to planetesimals. Next we review the late stage of accretion of terrestrial planets focusing on the Solar System. We discuss the numerical tools used to study terrestrial planet formation, the ingredients for such simulations and specific Solar System constraints on those simulations.  We present different views for the origin of the inner Solar System and possible ways to discriminate different models, concluding the section with a brief discussion of the long-term dynamical stability of the inner Solar System. Next we turn our attention toward terrestrial planet formation in the context of exoplanets. We discuss the orbital architecture of observed exoplanets and compositional constraints, the fate of terrestrial planets in systems with hot super-Earths and gas giant planets, and the dynamical evolution of exoplanet systems. Finally, we synthesize these different views and paint a picture of the Solar System in the context of exoplanets. 


\section{The early stages of planet formation}

Planet formation starts during star formation. The densest parts of a molecular cloud of gas and dust collapses due to its own gravity forming a protostar \citep{shuetal1987,mckeeostriker07,andreetal14}. Conservation of angular momentum turns the surrounding clump around the forming star into a circumstellar disk, the birthplace of planets \cite[e.g.][]{safronov72}. The primary empirical evidence that planets form according this paradigm is the geometry of our Solar System.  The planets' orbits are almost perfectly coplanar and orbit the sun in a common direction. In the very early stages (as for the interstellar medium) protoplanetary disks are 99\% gas and 1\% dust grains or ice particles \cite[e.g.][]{willianscieza11}. Despite the two orders of magnitude difference in abundance between these components, a disproportionately large part of planet formation studies have focused on the solid component. The gas component of the disk is much harder to observe because the gas only emits at specific wavelengths depending its composition and these emission lines also requires high resolution spectroscopy to be resolved \citep{najitaetal07,willianscieza11,dutreyetal04}. On the other hand, the solid counterpart has been used quite successfully as a disk tracer. Imaging of dust emission at infrared and submillimetre wavelengths and coronagraph imaging of scattered light at optical and near infra-red wavelengths have confirmed the distribution of dust in rotating, pressure supported disk-like structures \citep[e.g.][]{smithterrile84,beckwithetal90,odellwen94,hollandetal98,koerneretal98,schneideretal99}. Large far-infrared excesses also suggest that protoplanetary disk are not two-dimensional structures but rather possesses a vertical height  that increases with radius known as flaring \citep{kenyonhartmann87}. Recent observations from the ALMA have revealed never-before-seen details of planet forming-disks around young stars which shows carved ring-type structures \citep{vandermareletal13,isellaetal13,almaetal15,nomuraetal16,fedeleetal17}. These structures have been interpreted as signposts of planet formation in action \citep{dongetal15} or dust responding to the gas mainly via drag redistribution \citep[e.g.][]{takeuchiartymowicz01,flocketal15}.

 Solids orbit the star at the nominal Keplerian speed. Gas orbits the star at a slightly sub-Keplerian orbital velocity because protoplanetary disks are partially supported against the central star's gravity by a radial gradient in gas pressure. When the disk forms, dust is well mixed with the gas, but dust grains tend to settle into the disk mid-plane due the action of stellar gravity\citep[e.g.][]{weidenschilling80,nakagawaetal86}. Observations and simulations suggest that the disk itself falls into the central stars as consequence of radial angular momentum transport across the disk \citep[e.g.][]{papaloizoulin95,balbus03,dullemondetal07,armitage11}. As the disk loses mass by accretion into the star and also through photo-evaporation by ultraviolet and X-ray radiation \citep{gortihollenbach05} it eventually evolves from optically thick to optically thin \citep{alexanderetal14}. The lifetime of protoplanetary disks has been estimated from stellar spectroscopy of gas accretion on to the star, magnetospheric accretion models, and  infra-red excess measurements in the hot/inner parts of the disk to be shorter than a 10 Myr \citep{haischetal01,Hillenbrand08,mamajek09}. Dust grains absorb stellar photons and reradiate them at different wavelengths depending on its temperature  \citep[e.g.][]{willianscieza11}. Most observed disks in stellar clusters show no sign of near infra-red excess after 3 Myr \citep[e.g.][]{mamajek09}. The lifetimes of protoplanetary disks is a fundamental constraint on planet formation. 


It is convenient to divide planet formation into steps. Different physical and chemical process are at play at different stages of planetary growth, from dust grains to gas giants. For example, while the growth of micrometer-sized dust grains growth is driven by forces at the intermolecular level, the growth of km-sized and larger bodies is dominated by gravity. In addition, it is currently impossible to use numerical simulations to model the complete process of planet formation because of modeling and numerical limitations. In the following sections we briefly discuss each stage of planet formation. Namely, we discuss the growth from dust to pebbles, pebbles to planetesimals, planetesimals to planetary embryos, and planetary embryos to planets. 

\subsection{From dust to pebbles}

The solids available for planet formation are not uniform across a planet-forming disk.  At any given radius, only species with condensation/sublimation temperatures below the local temperature can exist as solids~\citep{grossmanlarimer74}. This leads to the concept of condensation or sublimation fronts, the best-known of which is the water ice snow line. Of course, as the disk cools in time, the locations of different condensation fronts sweep inward~\citep[e.g.][]{lecaretal06,dodsonrobinsonetal09,hasegawapudritz11,martinlivio12}.  Iron and silicates are abundant in the inner regions of the disk while the outer regions are rich in ice and other volatiles \citep{lodders03}. The first stage of planetary growth starts from the growth of micrometer sized dust/ice grains up to millimeter and cm-sized dust aggregates \citep[e.g.][]{lissauer93}. This view is well supported by disk observations and chondrite meteorites in our collections. Observations of young stellar objects at millimeter and centimeter wavelengths detect dust grains at these size ranges \citep{testietal03,wilneretal05,rodmannetal06,braueretal07,nattaetal07,riccietal10,testietal14,ansdeelletal17}. Yet, primitive meteorites are mainly composed of millimeter-sized silicate spheres known as chondrules \citep{shuetal01,scott07} cemented together by a fine-grained matrix of calcium aluminum rich inclusions (CAIs), unprocessed interstellar grains and primitive organics \citep{scotttaylor83,scottkrot14}. CAIs are thought to have been the first solids to condense in the protoplanetary disk, around 4.567 Gyr ago as estimated from the decay of short-lived radionuclides as  for example ${^{26}\rm {\rm Al}}$, which decays to $^{26}{\rm Mg}$ with a half-time of $\sim$0.7 Myr \citep[e.g.][]{bouvierwadhwa10,dauphaschaussidon11}. Chondrule formation may have started at the same time CAIs were forming and probably it last up to a few million years \citep{connellyetal12}. Chondrules may be splashes or impact jets produced during planetesimal collisions \citep[e.g.][]{asphaugetal11,johnsonetal15,wakita17,lichtenbergetal17} but their properties are more consistent with being rapidly-heated aggregates of mm-sized silicate dust grains \citep{deschconnolly02,morrisdesch10}.

Indeed, laboratory experiments and numerical simulations  suggest that the first stage of dust/ice growth is characterized by hit-and-stick collisions \citep[][]{blumwurm00,blumetal00,poppeetal00,guttleretal10,testietal14}.  The effects of electrostatic charges and magnetic material have been proposed as important for the interaction and growth of dust grains in protoplanetary disks \citep{dominiknubold02,okuzumi09} but  growth is probably mainly dictated by adhesive van der Wals forces  \citep{heimetal99,gundlachetal11}. Micrometer sized dust grains grow to larger fluffy porous aggregates by sticking \citep{blumwurm08} and  are eventually compacted by collisions to mm or cm-sized grains \citep{ormeletal07,zsometal10}. There is a general consensus that collisional growth of micrometer-sized dust grains is efficient up to mm to cm range in sufficiently dense regions of protoplanetary disks.

\subsection{From pebbles to Planetesimals}

Further growth of mm and cm-sized particles faces several obstacles \citep[e.g][]{testietal14}.  This subject is one of the most active current research areas in planet formation.
 
Numerical and laboratory experiments suggest that colliding millimeter and centimeter sized dust grains do not grow up to meter and kilometer sized bodie \citep{chokshietal93,dominiketielens97,gortietal15,krijtetal16}. Depending on particles' sizes and impact velocities, colliding dust particles and/or aggregates may bounce off of each other instead of growing \citep[e.g.][]{wadaetal09}.  This is known as the ``bouncing barrier''  \citep[e.g.][]{zsometal10,birnstieletal11,testietal14}. While the bouncing barrier poses a challenge to the growth of similar-size dust grains colliding at energetic speeds, it may not represent the end of dust growth. Energetic impacts between small projectiles and relatively larger targets can lead to mass transfer, helping to alleviate the bouncing barrier problem \citep{wurmetal05,teiserwurm09,kotheetal10,windmarketal12a,windmarketal12b}. However, even if mm or cm-size bodies can grow by pair-wise collisions \citep[e.g.][]{okuzumietal12} they face another challenge to reach the kilometer size scale. Because the gas' orbital speed is slightly sub-Keplerian, a particle on a Keplerian orbit feels a headwind that provides a drag force \citep{whipple72,adachietal76,weidenschilling77,haghighipourboss03}. Only sufficiently small particles that are strongly coupled to the gas do not drift as consequence of this aerodynamic effect. However, decimeter to meter-sized particles lie in a regime where they feel a very strong gas drag and spiral inwards very rapidly. A 1-meter size object at $\sim$1 AU in a typical disk falls toward the star in $\sim 100$ years \citep[e.g][]{weidenschilling77}. Particles of different sizes drift towards the star at different speeds. Large differential radial speed of these particles result in collisions at high speeds, leading to bouncing, fragmentation or erosion \citep{krijetetal15}. This problem is known as the  drift-fragmentation ``barrier''.

Another mechanism may explain how nature overcomes these barriers to produce macroscopic solid bodies. The idea is to bypass the critical size for rapid particle drift. If self-gravity is strong enough then planetesimals may form by gravitational collapse \citep{goldreichward73,youdinshu02,johansenetal14}. In order for mm or cm size particles to collapse they must be locally concentrated enough. Although these particles sediment to the disk mid-plane, the collapse of a dense mid-plane layer is generally prevented by turbulent diffusion \citep{weidenschilling80,cuzzietal08,johansenetal09}. Gravitational collapse only takes place if the local density is high enough to satisfy the following condition: $ 1 > (9 \Omega_c^2)/(4 \pi \rho_c G) $, where $\Omega_c$ is the Keplerian angular velocity at the clump location, G is the gravitational constant and $\rho_c$ is the mean bulk density of the potentially collapsing clump \citep[e.g.][]{johansenetal11,binneytremaine08,reinetal10,shichiang13}.  The latter expression is essentially the Toomre stability criterion for gravitational collapse \citep{toomre64} and it measures the interplay  between  stabilizing  rotation,  de-stabilizing self-gravity, and stabilizing  pressure of the clump \citep{binneytremaine08}. Different mechanisms may operate to concentrate particles locally in the gaseous protoplanetary disk at required levels for collapse \citep[see for example][]{balbushawley91,kretkelin07,lyraetal08a,braueretal08,lyraetal08b,lyraetal09,chambers10,drazkowskaetal13,squirehopkins17}.

Turbulent motion of the gas may also concentrate particles in eddies of rotating turbulent small gas structures \citep{cuzziweidenschilling06,chambers10}. Instabilities such as the vertical shear instability \citep{nelsonetal13,linyoudin15,barkerlatter15,umurhanetal16} and the baroclinic instability \citep{lyraklahr11,raettigetal13,bargeetal16,stollkley16} are also potential candidates for creating overdense regions in the disk \citep{johansenlambrechts17}. Drifting particles may also be trapped at localized higher pressure regions in the disks known as pressure bumps \citep{weidenschilling80}. In the vicinity of a pressure bump the pressure gradient is positive such that gas is accelerated, sometimes reaching super Keplerian speeds. Particles drifting inward are slowed down, halted or even reversed depending on the steepness of the local pressure gradient \cite[e.g.][]{haghighipourboss03}.  Pressure bumps may exist as a result of a sharp transition in the disc viscosity or due to the tidal perturbation from a sufficiently large planet  \citep[for a more detailed discussion see reviews by][]{chiangyoudin10,johansenetal14,johansenlambrechts17}. If the local density in solids increases by an order of magnitude at the pressure bump the level of turbulence may be significantly counterbalanced allowing planetesimal formation by gravitational collapse \citep{youdinshu02}. The local density may also be sufficiently increased by the delivery of slowly drifting small particles released during fragmentation/sublimation of larger  bodies as consequence of their entrance in hotter regions of the disk \citep{sirono11,idaguillot16}. This may be well the case of mm or cm-sided pebbles forming in the water and volatiles-rich outer regions of the disk and drifted inwards crossing the disk water iceline (a particular distance from the central star where water in the protoplanetary condenses as ice solid grains). If inward drifting pebbles cross the iceline their water-bulk component sublimates and small solid silicate/metals grains in their interior are released \citep{morbidellietal15}. 

The back-reaction friction force from the particles (mm or cm-sized) on the gas also plays an  aerodynamic role for concentrating particles locally in the disk \citep{youdingoodman05}. Because the gas is at sub-Keplerian speed, the back reaction of gas drag felt by a solid particle tends to accelerate the gas. If mm or cm-sized solid particles cluster sufficiently their collective back reaction becomes more pronounced and this reduces the speed at which the cluster drifts inwards. This allows that individual particles drifting at nominal drift speed from the outer regions of the disk to join the slowly inward drifting forming-cluster \citep{johansenyoudin07}. The successive addition of new members to the cluster increases the local density of solids  and further reduces the cluster's drift rate \citep{nakagawaetal86,johansenetal12}. If the local dust/gas ratio meets a threshold for starting the collapse, the cluster shrinks sufficiently due to energy dissipation by gas drag and collisions and the gravitational collapse may be successful in forming planetesimals in the range of sizes from $\sim$1 to $\sim$1000 Km \citep{johansenetal07,baistone10,simonetal16,schaferetal17,carreraetal17}. This gas drag assisted mechanism has been named ``streaming instability''. Streaming instabilities may be triggered preferentially outside the disk snowline where ice sticky particles form \citep{drazkowskadullemond14,armitageetal16,drazkowskaalibert17}.  Although many questions remain (see chapters by Andrews and Birnstiel), this scenario constitutes an appealing solution to how planetesimals form.

\subsection{From planetesimals to planetary embryos }

Once km-sized planetesimals have formed, their mutual gravity starts to play an important role. At this stage, planetesimals may grow in two different regimes, either by collisions between planetesimals or by accreting the remaining mm or cm-sized grains in the gas disk. We first discuss the scenario where planetesimals grow by mutual accretion. 

\subsubsection{Planetesimal-planetesimal growth} 

Let us assume that there exists a population of planetesimals with $N$ members and total mass $Nm$ at 1 AU.  These planetesimals have formed by a combination of the mechanisms described above, and are still embedded in the gaseous protoplanetary disk. The initial masses and physical radii of individual planetesimal are denoted $m$ and $R$, respectively.  Under mutual gravitational interaction these planetesimals evolve dynamically as they collide and scatter each other. Close encounters between planetesimals increase their random velocities by increasing their orbital inclinations and eccentricities. The random velocity $v_{rnd}$ represents the deviation of the planetesimal's velocity relative to a Keplerian circular and planar orbit at its location:
\begin{equation}
v_{rnd} = (\frac{5}{8}e^2 + i^2)^{1/2} v_{k},
\end{equation}
where $e$  and $i$ are the planetesimal's orbital eccentricities and inclinations \citep{safronov72,greeenbergetal91}. $v_k=\sqrt{\frac{GM\odot}{r}}$ is the planetesimal Keplerian velocity, where $G$ is the gravitational constant, $M_\odot$ is the mass of the central star, and $r$ is the planetesimal's orbital radius. 

Relative velocities among planetesimals determine how fast they grow. What is important is the ratio of their random velocities to the local Keplerian velocity and the differential shear across the  Hill radii of interacting bodies. The Hill radius $R_H$ of a planetesimal orbiting a star with mass $M_\odot$ is defined as
\begin{equation}
R_{H} = a\left( \frac{m}{3M_\odot}\right)^{1/3},
\end{equation}
where $a$ and $m$ are the planetesimal's semi-major axis and mass, respectively. 

Gravitational scattering among planetesimals is in the shear-dominated regime when $v_{rnd} < R_H \Omega$ and in the dispersion-dominated regime when $v_{rnd} > R_H \Omega$, where $\Omega$ is the local Keplerian frequency. In the shear-dominated regime, planetesimals during a close-encounter spend some time close to each other. In this case, gravity may sufficiently  deflect their orbits and ``focus'' their trajectories towards each other. This effect increases their collision probability and speeds up growth.  However, in the dispersion-dominated regime interacting planetesimals spend little time close to each other because of their high relative speeds. Gravitational focusing is less efficient, which decreases their collision probabilities increases the accretion timescales. 
 
Random velocities of planetesimals are also damped by dissipative effects. These include physical collisions \citep{goldreichtremaine78}, gas drag \citep{adachietal76} and gas dynamical friction \citep{grishinperets15}. The balance between excitation and damping changes over time as the disk dissipates and planetesimals grow.

Planetesimals grow to planetary embryos to planets in three different regimes. In the so called ``Runaway growth'' regime larger planetesimals grow faster than smaller ones forming the first planetary embryos.   In the ``orderly growth'' regime planetary  embryos enters in the regime where they grow roughly at the same rate, eventually forming planets. The ``Oligarchic growth'' regime is an intermediate scenario. 

One can write the accretion rate of a planetesimal with mass $m$ as \citep{safronov72,idanakazawa89,greenzweiglissauer90,rafikov03}
\begin{equation}
\frac{dm}{dt} \simeq \pi {R}^2 \Omega \Sigma \frac{v_{rnd}}{v_{rnd,z}} \left(1+\frac{2Gm}{R v_{rnd}^2} \right),
\end{equation}
where  $R$ is the planetesimal physical radius, $\Omega$ is the planetesimal Keplerian frequency, $\Sigma$ is the local planetesimal surface density, and $v_{rnd,z} \neq 0$ is the averaged planetesimal velocity in the vertical component.  In Equation 3, the term $\frac{2Gm}{R v_{rnd}^2}$ represents the gravitational focusing and it is simply the ratio of the scape velocity from the planetesimal surface to the relative velocity of the interacting planetesimals squared. Also in Eq. 3, $\Sigma \approx N  \overline{m} H$ where N is the is the surface number density of planetesimals, $\overline{m}$ is the averaged masses of planetesimals and $H \approx v_{rnd,z}/\Omega$ represents the vertical thickness of the region where planetesimals are in the disk. Eq. 3 neglects bouncing, fragmentation or erosion during collisions.

So-called ``dynamical friction'' plays an important role during accretion. Because of angular momentum transfer, during close-encounters with smaller bodies the largest planetesimals tend to have their orbital eccentricities and inclinations damped. As result, their random velocities decrease, favoring strong gravitational focusing for those large bodies. If (1) gravitational focusing is large (or planetesimal are small); (2) planetesimals' random velocities are roughly independent of their masses ($v_{rnd}$ and $v_{rnd,z}$); and (3) growth does not strongly affect the surface density of planetesimals, then the Equation 3 may be simplified as \citep{wetherillstewart89,kokuboida96}
\begin{equation}
\frac{1}{m}\frac{dm}{dt} \approx t_{grow,run} \approx  \Sigma \frac{1}{v_{rnd,z}}\frac{{m}^{1/3}}{v_{rnd}} \approx {m}^{1/3}.
\end{equation}
Note that the $m^{1/3}$ term arise simply from ${R\approx m^{1/3}}$. In this regime the accretion rate increases for more massive planetesimals. If $m_1$ and $m_2$ are the masses of two planetesimals in the disk where $m_1 > m_2$ then in the runaway growth $\frac{d(m_1/m_2)}{dt} > 0$. This is the so called ``Runaway growth'' regime of planetesimal accretion.

Of course, the runaway growth regime must include its fair share of non-accretionary collisions \citep{leinhardtrichardson05b,stewartleinhardt09}. If small fragments are produced as the outcome of collisions during the runaway growth they may have a second chance to be accreted by other planetesimals if the gaseous disk has not yet dissipated. Small fragments' orbits are quickly damped by gas drag which in turn increases their gravitational focusing and the probably of being consumed by larger planetesimals \citep{wetherillstewart93,rafikov04}. However, this is only possible  if small fragments do not drift inwards too quickly too be  accreted \citep{kenyonluu99,inabaetal03,kobayashietal10} .

The first population of planetary embryos emerges from the largest planetesimals growing by runaway accretion \citep{kokuboida96,ormeletal10}. Gravitational perturbations from emerging embryos then start to dominate over the influence of smaller planetesimals and the growth regime changes. Embryos increase the random velocities of planetesimals and change the local surface density of planetesimals \citep{tanakaida97}. The local velocity dispersion thus depends on the embryo mass. This causes a transition to a new regime in which the embryo/planetesimal growth rate takes the form
\citep{wetherillstewart89,kokuboida96}
\begin{equation}
\frac{1}{m}\frac{dm}{dt} \approx  t_{grow,orl} \approx  \Sigma \frac{1}{v_{rnd,z}}\frac{{m}^{1/3}}{v_{rnd}}
\end{equation}
This mode of growth is denominated  ``Oligarchic'' and it is a phase between the runaway and orderly growth regimes.  The oligarchic growth rate depends on the scaling of $\Sigma$, $v_{rnd}$ and $v_{rnd,z}$ \citep{rafikov03}. In this regime planetary embryos still grow faster than planetesimals. Thus, if $m_1$ represents the mass of a planetary embryo and  $m_2$ the mass of a planetesimal, $\frac{d(m_1/m_2)}{dt} > 0$. However, small planetary embryos ($m_2$) may grow faster than larger ones $\frac{d(m_1/m_2)}{dt} < 0$. According to \citep{idamakino93} the transition from the runaway to the oligarchic regime occurs  when the masses in planetary embryos is about 2 times larger than the counterpart in planetesimals \citep[but see ][for a different criterion]{ormeletal10b}.

The oligarchic regime generates a bi-modal population of planetary embryos and planetesimals. The total mass in embryos is smaller than the total mass in planetesimals. Embryos growing in the planetesimal sea delimit their own regions of gravitational influence. Planetary embryos are typically spaced from each other by 5-10 mutual Hill radii \citep{kokuboida95,kokuboida98,kokuboida00,ormeletal10}, where the mutual Hill radius of two embryos with masses $m_i$ and $m_j$ and semi-major axes $a_i$ and $a_j$ is defined as
\begin{equation}
R_{H,i,j} = \frac{a_i+a_j}{2}\left( \frac{m_i+m_j}{3M_\odot}\right)^{1/3}.
\end{equation}
If two embryos come closer than a few mutual Hill radii, they scatter each other and increase their orbital separation to again be larger than $\sim$5$R_{H,i,j}$. After embryo-embryo scattering events, dynamical friction (or gas drag) tends to circularize their orbits and damp  orbital inclinations if there is enough mass in planetesimals or gas in the embryo's vicinity. 

\begin{figure}
\centering
\includegraphics[scale=.15]{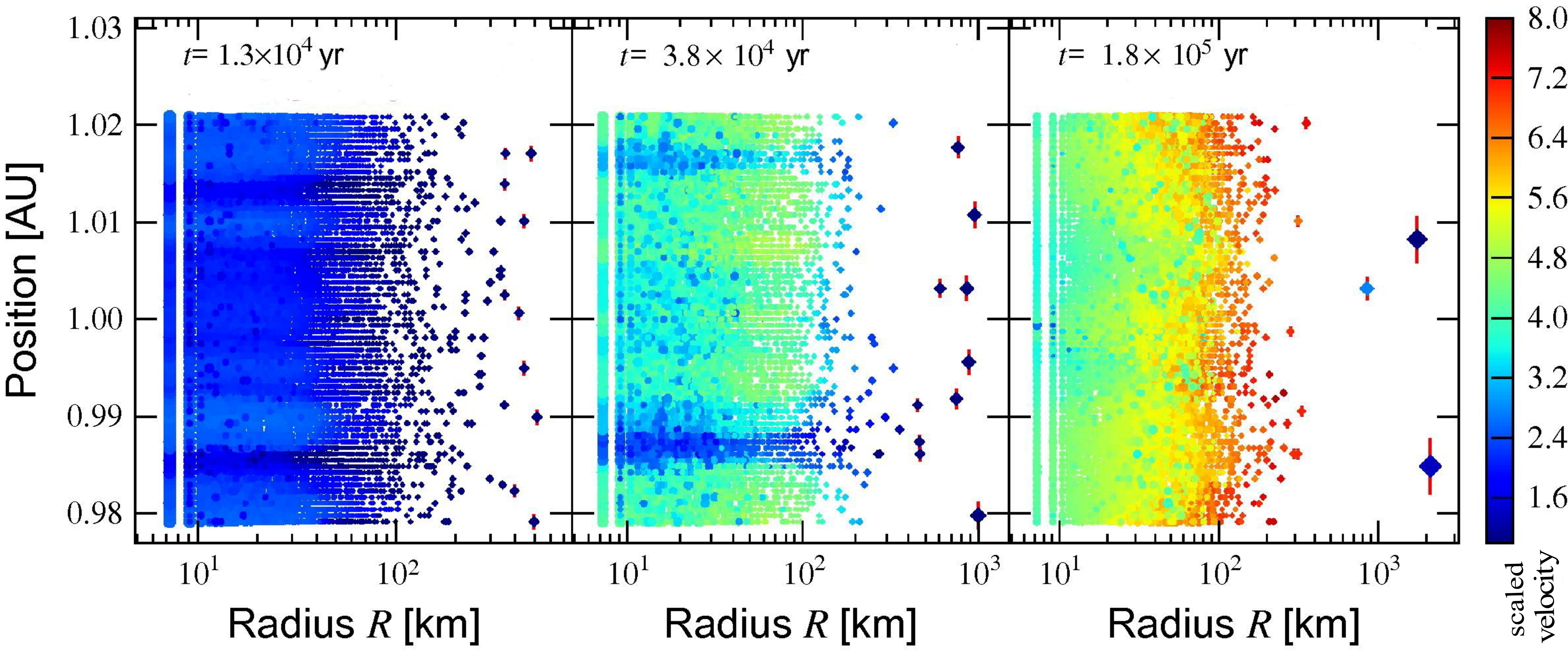}
\caption{ Planetesimal and planetary embryo growth during the oligarchic regime. Dots represent swarms of planetesimals which are dynamically evolved as a single entity. The dot size   scales  as $m_s^{1/3}$ where $m_s$ is the swarm  total mass. Individual large bodies are shown as diamonds. The color represents the scaled random velocities of the bodies. The red bar intercepting the largest bodies in the system represents the respective  size of their Hill radius.   Figure adapted from \citep{ormeletal10}}
\label{fig:1}       
\end{figure}

Figure \ref{fig:1} shows the growth of planetesimals and embryos during the oligarchic growth regime \citep{ormeletal10}. Three snapshots in the system's are shown. The code uses a Monte Carlo method to calculate the collisional and dynamical evolution of the system  \citep{ormeletal10} and tracer particles to represent swarm of small planetesimals. This approximation reduces the number of iterations needed to solve the system's dynamical evolution while maintaining to some degree the individual nature of particles. Figure  \ref{fig:1} shows that after 0.18 Myr two prominent planetary embryos with physical radius larger than 1000 km emerge in the disk with mutual separation of a few Hill radii.

\subsubsection{Pebble accretion: from planetesimals to planets}

If planetesimals form early they may accrete dust grains drifting within the disk. The existence of such grains in planet-forming disks has been observationally confirmed \citep[e.g.][]{testietal14}. The accretion of mm or cm-sized grains by a more massive body is commonly known as pebble accretion \citep{johansenlacerda10,ormelklahr10,lambrechtsjohansen12,johansenetal15,xuetal17}. 

Pebble accretion can be much faster than planetesimal accretion and may solve some long-standing problems.  In the ``core accretion'' scenario, a gas giant planet forms by the accretion of gas onto a  $\sim$10$M_\oplus$ masses~\citep{mizuno80,pollacketal96}. To become a gas giant a core must form before the gaseous disk disperses otherwise. However, simulations of planetesimal growth struggle to grow the cores of Jupiter or Saturn within a typical disk lifetime~\citep[e.g.][]{thommesetal03,levisonetal10}.

Millimeter- to centimeter-sized pebbles spiral inwards rapidly due to gas drag \citep{adachietal76,johansenetal15}. The dynamical behavior  of a single drifting pebble approaching a planetesimal from a relatively more distant orbit is determined by a dramatic competition between gas drag and its mutual gravitational interaction with the larger body. It is assumed that the planetesimal is  sufficiently small and that it does not disturb the background gas disk structure  (e.g. gas disk velocity and density). Two end-members outcomes are possible during the encounter pebble-planetesimal. The pebble may either cross the planetesimal orbit without being accreted or may have its original orbit sufficiently deflected to be accreted.

Pebble accretion onto sufficiently large bodies can be extremely fast. However, planetesimals (or planetary embryos) cannot grow indefinitely even if the pebble flux is high. As an embryo grows, it gravitationally perturbs the structure of the gaseous disk. Eventually, the growing body opens a shallow gap and creates a local pressure bump outside its orbit. If the pressure bump is sufficiently high, particles entering the bump are accelerated by the gas and stop drifting inwards. At the ``Pebble isolation mass'' $M_{iso}$ an embryo or planet stops accreting pebbles,
\begin{equation}
M_{iso} = 20\left( \frac{H_{gas}/a_p}{0.05}\right)^3 M_\oplus,
\end{equation}
where $H_{gas}$ is the gas disk scale height \citep{lambrechtsetal14,morbidellinesvorny12}. It is worth noting that two recent studies have proposed that pebbles may be partially or even fully evaporated/destroyed before they can reach the accreting core. This effect may become important before the core reach isolation mass \citep{alibert17,brouwersetal17}. Further study is needed to understand exactly how this effect limits embryo growth by pebble accretion.

\subsection{From planetary embryos to planets} 

If embryos grow to be $\sim$100 times more massive than individual planetesimals the random velocities of planetesimals scale as $v_{rnd} \approx m^{1/3}$.  Also, it is possible that by this stage the gaseous protoplanetary has dissipated, removing the dissipative mechanisms of gas drag and gas dynamical friction. In such an environment, gravitational focusing becomes negligible  and this drastically lengthens the accretion timescale. Given that $v_{rnd} \approx v_{rnd,z}$ \citep{rafikov03} Eq 3 takes the form
\begin{equation}
\frac{1}{m}\frac{dm}{dt} \approx t_{grow,ord} \approx  \Sigma \frac{v_{rnd}}{v_{rnd,z}}{m}^{-1/3} \approx \Sigma {m}^{-1/3}
\end{equation}
In this mode of growth, termed  ``orderly growth'' or ``late stage accretion''  $\Sigma$ decreases because massive embryos accrete or scatter nearby planetesimals and open large gaps in the disk \citep{tanakaida97}. At this stage, most of the mass is carried by embryos rather than planetesimals. All planetary embryos grow at a similar rate and their mass ratios tend towards unit. Considering $m_1$ and $m_2$ the masses of two planetary embryos in this regime, where $m_1 > m2$, this scenario results $\frac{d(m_1/m_2)}{dt} \approx 0$.

Orderly growth is marked by violent giant collisions between planetary embryos, scattering and ejection of planetesimals and planetary embryos.  The system evolves chaotically and the planetesimals population decreases drastically. Assuming that 50\% of the total mass in planetesimals is carried by embryos~\citep{kenyonbromley06}, the  mass of an embryo at the start of orderly growth is $M_{ord}  = \int_{r-\Delta_r/2}^{r+\Delta r/2} 2 \pi r'  \Sigma(r')/2  dr' \simeq \pi r \Delta r \Sigma$, where  $\Delta r$ corresponds to the width of the feeding zone of the embryo  and r is the  planetary embryo heliocentric distance \citep{lissauer87}. The size of the feeding zone of a embryo typically ranges between a few to 10$R_{H}$. In our Solar System, embryos probably reached  masses between those of the moon and Mars in the terrestrial region. This is consistent with the typical masses of planetary embryos produced in  oligarchic regime \citep{kokuboida00,chambers01}. The oligarchic growth timescale of a Mars-mass embryo with bulk density $\rho_p=$ 3$g/cm^3$ orbiting at $a_p=1 AU$ and interacting with planetesimals with average random velocities  $v_{rnd}=v_{rnd,z}=0.02v_k$ is about 0.37 Myr for a local surface density in planetesimals of $\Sigma=10 g/cm^2$. The growth timescale of this same planetary embryo in the orderly regime  is about two orders of magnitude longer (23 Myr). This highlights the dramatic role of gravitational focusing.

\subsection{Methods and Numerical tools}

Studies of the runaway and oligarchic regimes have been also conducted in  different fronts. This includes N-body simulations \citep{idamakino93,aarsethetal93, kokuboida96,kokuboida98,kokuboida00,richardsonetal00,thommesetal03,barnesetal09}, analytical/semi-analytical calculations based on statistical algorithms \citep{greenbergetal78,wetherillstewart89,rafikov03,rafikov03b,rafikov03c,goldreichetal04,kenybromley04,idalin04,chambers06,morbidellietal09,schlichtingsari11,schlichtingetal13}, hybrid statistical/N-body (or N-body coagulation) codes which incorporates the two latter approaches \citep{spauteetal91,weidenschillingetal97,ormeletal10,bromleykenyon11,glaschkeetal14}, and finally the more recently developed hybrid particle-based algorithms \citep{levisonetal12,morishimaetal15,morishima17}.

Statistical or semi-analytical coagulation methods model the dynamics and collisions of planetesimals in a ``particle-in-a-box approximation'' \citep{greenbergetal78}. This method is based on the kinetic theory of gases and it neglects the individual nature of the particles but rather uses distribution functions to describe the planetesimals orbits. This approach has been  used to model the early stages of planet formation when the number of planetary objects is large ($>>10^3$). While the statistical calculations give a statistical sense of the dynamics of a large population of gravitationally interacting objects they invoke a series of approximations which are only valid at local length scales in the protoplanetary disk \citep{goldreichetal04}. The necessity of including non-gravitational effects and collisional evolution typically leads to approaches that are not self-consistent \citep{leinhardt08}.

Direct N-body numerical simulations are typically far more precise the simple coagulation approaches but cannot handle more than a few thousand self-interacting bodies for long integration times (e.g. $\sim10^8-10^9$ yr) without reaching prohibitively long computational times. Each tool may be suitable to model different stages of planet formation. While studies of the early stages of planet formation are mostly conducted using analytical and statistical tools, the intermediate and late stage of accretion of planets are typically approached by direct N-body integration \citep{lecaraarseth86,beaugeaarseth90,chambers01,kominamiida04}. Hybrid statistical algorithms explore the main advantages of coagulation algorithms and N-body codes \citep{kenyonbromley06}. They can be used to follow the transition between different growth regimes of planet formation as for example planet formation from the runaway to the orderly phase.

There are several  numerical N-body integration packages to model planetary formation and dynamics  available as Mercury \citep{chambers99}, Symba \citep{duncanetal98}, Rebound \citep{reintamayo15,reinspiegel15}, Genga \citep{grimmstadel14}, HNBody \citep{rauchhamilton02}, etc . Among them, probably the most popular are Mercury (publicly available) and Symba. These  codes are built on symplectic algorithms which divide the full Hamiltonian of the problem in a part describing the keplerian motion and a part due to the mutual gravitational interaction of bodies in the system \citep{wisdomholman91}. Symplectic algorithms are conveniently applied to  systems where most of the total mass is carried by a single body, and they  allow long-term numerical integrations without the propagation and accumulation of errors \citep{sahatremaine94}. However, pure symplectic algorithms require a fixed integration step-size.  Nevertheless, planet formation simulations present close-encounters and collisions among planetary objects which require sufficiently small timesteps to be properly resolved \citep{chambers99}. Using a symplectic algorithm with a sufficiently small timestep to resolve planetary encounters destroys its speed advantage compared to traditional algorithms. The symplectic algorithms in Mercury and Symba are adapted to overcome this issue. Symba divides the full Hamiltonian of the problem and uses a different step-size to each part \citep{duncanetal98}. Mercury detects a close-encounter and solve the orbits of bodies involved in the approach with a different numerical integrator with  self-adaptive timestep \citep{chambers99} while the remaining terms are solved symplectically.

Particle-based algorithms have been also developed. These are hybrid algorithms in the sense they combine N-body direct integration with  super-particle approximation \citep{levisonetal12,morishimaetal15}. These codes can integrate a larger number of small particles represented by a small number of tracer particles. The tracers interact with the larger bodies thought a N-body scheme. Tracer-tracer interactions are solved using statistical routines modelling stirring, dynamical friction and collisional evolution. The LIPAD code \citep{levisonetal12} has been used, for example, to model the formation of terrestrial planets in the Solar System from a larger number of planetesimals \citep{walshlevison16} and also in  simulations of terrestrial formation including pebble accretion \citep{levisonetal15pnas}. 

\section{Late stage accretion of terrestrial planets in the Solar System}

In this section we discuss models of the late stage accretion of terrestrial planets in our own Solar System. We first discuss the constraints on these models, then we present different current scenarios that can match these constraints.  Finally, we discuss strategies to distinguish or falsify these models.

\subsection{Solar System Constraints}



\subsubsection{Planetary masses, orbits and number of planets}

The masses and orbits of the terrestrial planets are the strongest constraints for formation models. The orbits of the terrestrial planets have modest eccentricities and inclinations. The Angular Momentum Deficit (AMD) measures the level of dynamical excitation of a planetary system \citep{laskar97,chambers01}. The AMD of a planetary system measures the fraction of angular momentum missing compared to a system where the planets have the same semi-major axes but circular and coplanar orbits. The AMD is a diagnostic of how well simulated terrestrial planetary systems match the real terrestrial planets' level of dynamical excitation. The AMD is defined as
 \begin{equation}
{\rm AMD}= \frac{{\sum_{j=1}^N}\Big[ m_j \sqrt{a_j} \> \Big( 1 - \cos i_j \> \sqrt{1-{e_j}^2}\> \Big)\Big]} 
{{\sum_{j=1}^N}\> m_j \sqrt{a_j}}.
\end{equation},
where $m_j$ and $a_j$ are the mass and semi-major axis of each planet $j$, $N$ is the number of planets in the system, and ${\rm e_j}$ and ${\rm i_j}$ are the orbital eccentricity and inclination of each planet ${\rm j}$. The terrestrial planets' AMD is 0.0018.  

Another useful metric is the Radial Mass Concentration (RMC)~\citep{chambers01}, a measure of a planetary's system's degree of radial concentration. Earth and Venus contain more than 90\% of the terrestrial planets' total mass in a narrow region between 0.7 and 1 AU. The RMC is defined as : 
\begin{equation} 
{\rm RMC} = {\rm Max}\left(\frac{{\sum_{j=1}^N}\> m_j}{{\sum_{j=1}^N}\> m_j\big[\log_{10}\left(a/a_j\right)\big]^2}\right). 
\end{equation}
Higher values indicate more concentrated systems. The inner Solar System RMC is 89.9.

Exactly what cutoffs should be used when applying these metrics is somewhat arbitrary.  For success, studies often require the AMD and RMC to be matched to within a factor of 1.3-2 depending on the situation \citep[e.g.][]{raymondetal09,izidoroetal14}. Of course, viable Solar System analogs should have between 3 and 5 planets with semi-major axes between $\sim$0.3 and $\sim$1.8 AU.

\subsubsection{Timing of planet formation}

Given their large gas contents \citep{wetherill90,lissauer93,guillotetal04}, the giant planets are constrained to have formed prior to the dispersal of the gaseous protoplanetary disk \citep{bodenheimerpollack86,pollacketal96,alibertetal05}. Gas giants were probably fully formed before the end of terrestrial planet accretion. Virtually all models of terrestrial planet formation agree that giant planets play a critical role shaping the terrestrial planets \citep[e.g.][]{wetherill78,wetherill86,chamberswetherill98,agnoretal99,morbidellietal00,chambers01,raymondetal06,obrienetal06,lykawakaito13,raymondetal14,izidoroetal14,
fischerciesla14,lykawakaito17}. 
However, different giant planet configuration has been used to model the formation of the terrestrial planets. We will return to this issue later in this Section.

Numerical simulations and radiometric dating techniques agree that the last giant impact on Earth took place between $\sim$30 and $\sim$150 Myr after CAIs' formation \citep{yinetal02,jacobsen05,toubouletal07,allegreetal08,hallidayetal08,kleineetal09,jacobsonetal14}. Even if this event took place on the early branch of this interval (e.g. around 30 Myr), this is very likely after the nebula gas dispersal \citep{briceetal01,mamajek09} and after the giant planets were fully formed. Nevertheless, Mars probably is much older than the Earth. Radiometric dating of Martian meteorites using Hafnium-Tungsten (Hf-W) isotopes indicates that Mars reached about half of its current mass during the first 2 Myr after CAI formation \citep{duaphaspourmand11}. Given that Mars was fully formed by 10 Myr after CAIs \citep{nimmokleine07}, it may be as old as our gas giants. Venus and Mercury meteorites has not been identified in meteorites collections which makes their ages unconstrained.

\subsubsection{The Asteroid belt}

\begin{figure}
\centering
\includegraphics[scale=.5]{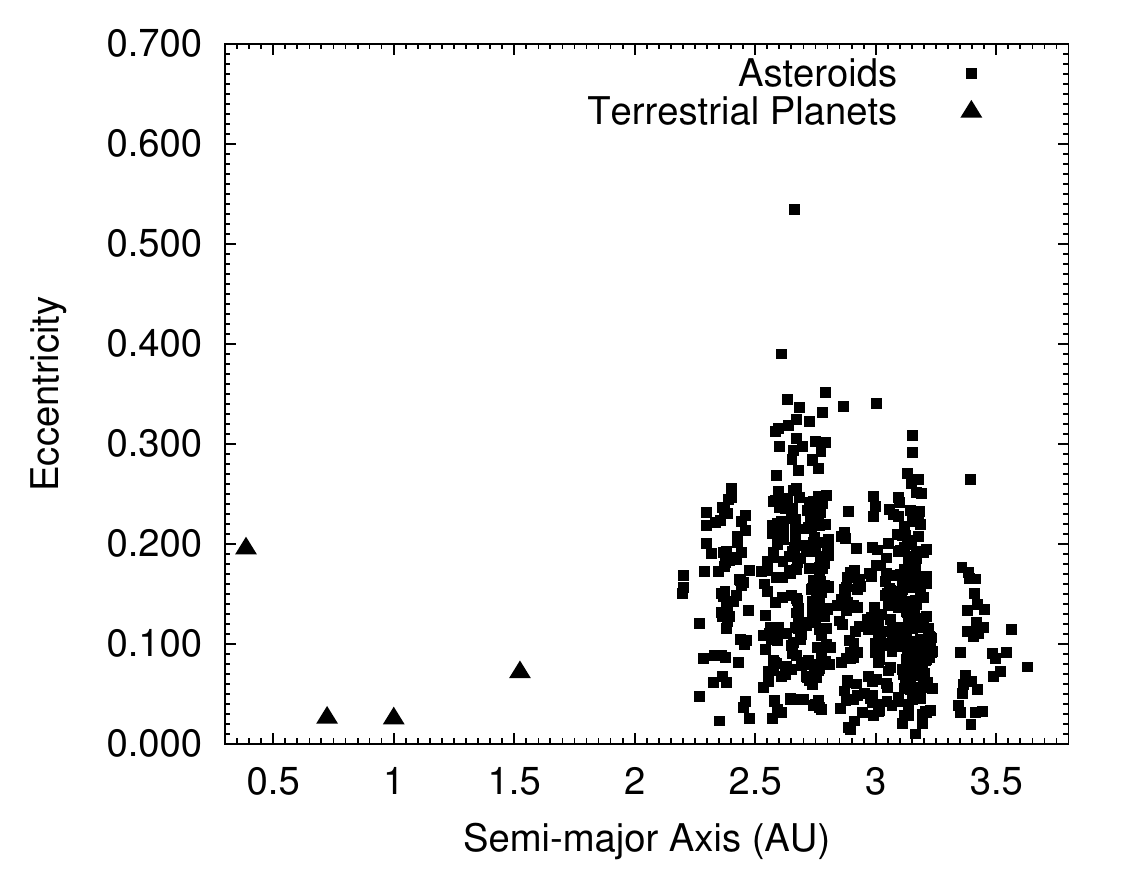}
\includegraphics[scale=.5]{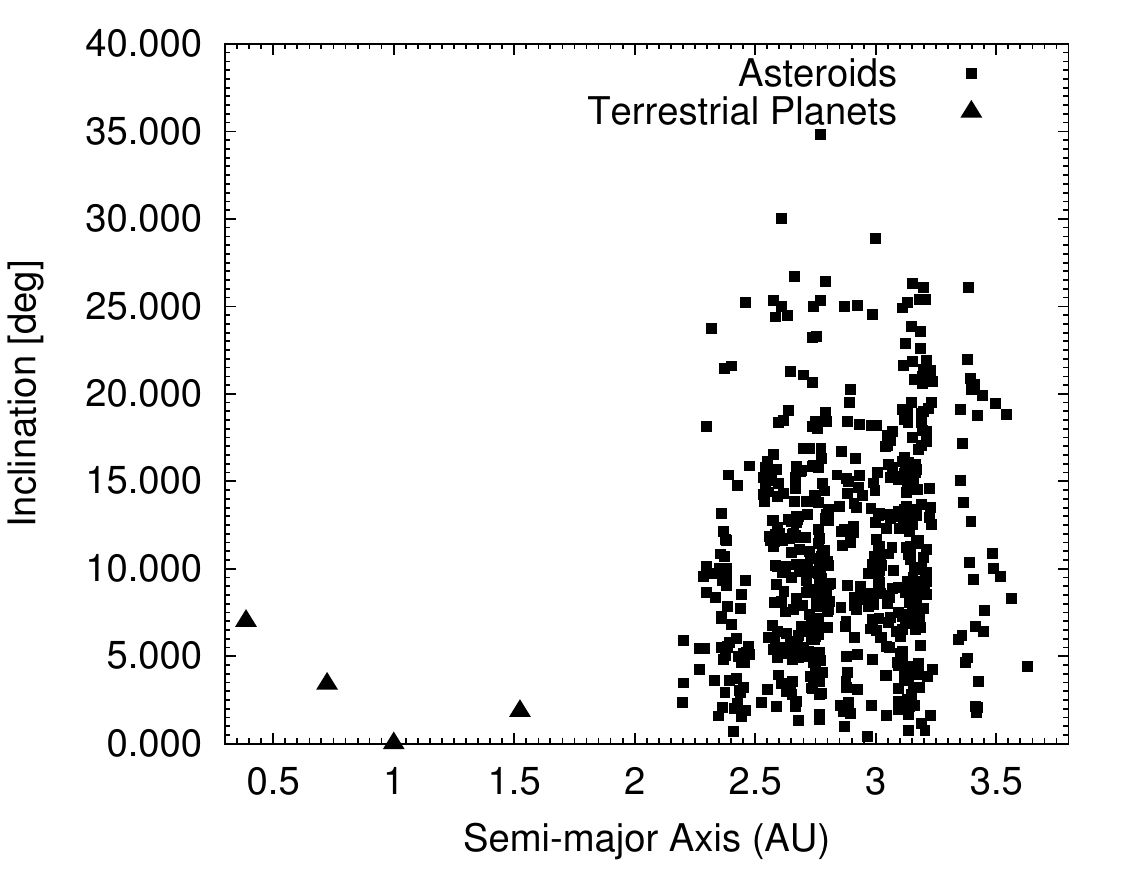}
\caption{Orbital architecture of the inner Solar System. In both panels, planets are shown as triangles and asteroids as squares. The left-hand panel show a diagram semi-major axis versus orbital inclination. The right-hand panel shows orbital semi-major axis versus orbital inclination. Only asteroids larger than 50 Km are shown.}
\label{fig:3}       
\end{figure}

Terrestrial and giant planets in our Solar System are physically separated by the asteroid belt. Unlike the reasonably circular and coplanar orbits of the planets, the asteroids are dynamically excited. Asteroid eccentricities ranges from 0 to $\sim$0.4 and their orbital inclinations between 0 and $\sim$ 25 degrees (Figure \ref{fig:3}). The asteroid belt is also mass depleted \citep[e.g.][]{petitetal01,petitetal02,morbidellietal15b}. The total mass in the terrestrial planets is about 2$M_\oplus$. Mercury and Mars' semimajor axes are about 0.38 AU and 1.5 AU, respectively. The inner edge of the main belt is at about 1.8 AU while the outer edge at 3.2 AU. If we could dispose all main belt asteroids and planets in a common plane the total area occupied by the asteroids' orbits is at least 3 times larger than that occupied by the terrestrial planets together. Yet the main asteroid belt region contains only roughly $5\times 10^{-4}\mearth$ \citep{gradietedesco82,demeocarry13,demeocarry14}. Ceres is most massive object in today's belt. Yet, it is very unlikely that the asteroid belt hosted in the past a planet much more massive than that (e.g. Mars).  Such massive object would have sculpted large gaps in the asteroid belt and such gaps are not observed today \citep{raymondetal09}.

The asteroid belt is chemically segregated \citep[e.g.][]{demeoetal15}. The inner region is mostly populated by silicaceous asteroids (S-type) while the outer region is dominated by Carbonaceous chondrites  asteroids (C-type). A variety of taxonomic type of asteroids exist in the belt but S and C-type are the dominant populations \citep{demeocarry14}. C-type asteroids are dark asteroids because they contain a large amount of carbon and hydrated minerals. S-type asteroids are moderately bright and are most composed of iron and silicate material \citep{gradietedesco82,demeocarry14}.

\subsubsection{Water on Earth and other terrestrial planets}

The amount of water on Earth is debated \citep[e.g.][]{drakecampins06}. Estimates suggest that the total Earth water content is between $\sim$1.5 and $\sim$10-40 Earth oceans \citep{lecuyeretal98,marty12,halliday13}, where 1 Earth ocean is $1.4\times 10^{24}g$.  A major part of this water is stranded in the Earth's mantle. More water may exist on Earth's core but the true amount is uncertain \citep{nomuraetal14,badroetal14}. Interestingly, the Earth contains more water than would be expected from the radial water gradient across the Solar System \cite[see the recent review by][]{obrienetal18}. Asteroids, mainly those in the outer part of the belt (e.g., semi-major axis larger than 2.5 AU), transneptunian objects and comets are very water rich, with concentration of up to 20\% water-mass fraction. However, asteroids belonging to different taxonomic types in the inner region of the asteroid belt (e.g. enstatite chondrites)  are typically drier than Earth \citep[e.g., see review by][]{morbidellietal12}. If the Earth accreted mainly from rocky material exposed to relatively higher temperatures in the protoplanetary disk than that material that accreted asteroids at larger distances it is reasonable to expect that the Earth  should be at least as reduced in  volatiles and water as  the innermost asteroids. Thus, given the larger amount of water on Earth it is believed that one or more mechanisms contributed delivering a major part of its water \citep[e.g.][]{morbidellietal00,raymondetal04,raymondetal07,izidoroetal13,obrienetal14,raymondizidoro17a}. 

There is evidence for water on Mercury \citep{lawrenceetal13,ekeetal16}. The high D/H ratio in Venus' atmosphere suggests that in the past the planet had a larger amount of water that escaped to space \citep{donahueetal1982,kastingpollack83,grinspoon93}. The high D/H of Mars' atmosphere and isotopic analysis of martian meteorites also suggest some of its primordial water was lost to space \citep[e.g.][]{owenetal88,kurokawaetal14}. Geomorphological features on Mars indicate the planet had ancient oceans and a lot of water may be hidden below the surface \citep{bakeretal91}. All this evidence supports the idea that a significant amount of water was delivered to the inner Solar System. 

Isotopic ratios are very useful to discriminate water sources. Carbonaceous chondrite meteorites are associated with C-type asteroids in the belt and their hydrogen  and nitrogen isotopic ratios --D/H and ${\rm ^{15}N/ ^{14}N}$-- match those of Earth \citep{martyreika06,marty12}. The D/H of the solar nebula is generally inferred from the Jupiter's atmosphere and it is estimated to be by a factor of $\sim$5-10 lower than carbonaceous chondrites.  Water with a D/H ratio similar to the solar value  has been found in Earth's deep mantle \citep{hallisetal15} but in order for Earth's water to have a primarily nebular origin one must invoke a mechanism to increase the D/H ratio of Earth's water over the planet's history. In principle this could be achieved if  the Earth had a massive primordial hydrogen-rich atmosphere that efficiently escaped to the space over a billion year timescale  due to very intense UV flux \citep{gendaikoma06,gendaikoma08}. However, the  solar ${\rm ^{15}N/ ^{14}N}$ ratio also does not match that of Earth \citep{marty12}.

Comets present a wide range of D/H ratios, which vary from terrestrial-like to several times higher \citep{alexanderetal12}. Nevertheless,  elemental abundances and mass balance calculations based on $^{36}$Ar suggest it is unlikely that comets contributed with more than a few percent of Earth's water \citep{martyetal16}, but this same analysis suggest they probably contributed nobles gases to Earth's atmosphere \citep{martyetal16,aviceetal17}. Carbonaceous chondrites remain the best candidates for delivering water to Earth \citep{alexanderetal12}. The much higher D/H ratios of Venus and Mars  probably do not represent their primordial values and their water origin remain unconstrained. However, any process delivering water to Earth would invariably also deliver water to the other terrestrial planets \cite[e.g.][]{morbidellietal00,raymondizidoro17a}.

\subsubsection{Giant planet orbits and evolution}

A fundamental dynamical constraint on terrestrial planet formation models is the orbital architecture of the gas giants. However, the giant planets' orbits were not necessarily the same at the time the terrestrial planet were forming. Hydrodynamical simulations show that the giant planets probably migrated during the gas disk phase.  The most likely outcome of migration is a chain of mean motion resonances among the giant planets     \citep{massetsnellgrove01,morbidellicrida07,dangelomarzari12,pierensetal14}. There is a growing consensus that the giant planets orbits evolved from a more compact configuration to  their current orbits through a dynamical instability  \citep{fernandezip84,hahnmalhotra99,tsiganisetal05,nesvornymorbidelli12}.  However, two different views exist on the timing of that instability/migration-phase \citep[for a more detailed discussion see][]{morbidellietal18}.  In one  view the gas giants reached their current orbits early in the Solar System history, probably before all the terrestrial planets were fully formed and likely a few tens of million years after CAIs \citep{kaibchambers16,toliouetal16,deiennoetal17}. In the second view, the giant planets' current orbits were only reached much later, probably 400-700 Myr after CAIs formed, coinciding with the so-called 'late heavy bombardment' \citep{gomesetal05,levisonetal11,deiennoetal17}.


If the giant planets only reached their current orbits very late, then they were likely in a low-eccentricity resonant configuration during terrestrial accretion \citep{raymondetal06,obrienetal06,raymondetal09,izidoroetal14,izidoroetal15,izidoroetal16}.  On the other hand, if the giant planets reached their current orbits early, with an instability that happened shortly after the dissipation of the gaseous disk, then the giant planets' orbits during accretion would be close to their present-day orbits. The orbital period ratio of Jupiter and Saturn today is $P_s/P_J \simeq 2.48 $, and their orbits are slightly eccentric and inclined (the importance of this issue will be justified later).

\subsection{Solar System Terrestrial Planet formation Models}

Simulations of late stage accretion of the terrestrial planets typically start from a population of already-formed planetesimals and Moon- to Mars-mass planetary embryos. This scenario is consistent with models of the runaway and oligarchic growth regimes \citep{kokuboida96,kokuboida98,kokuboida00,chambers06,ormeletal10,ormeletal10b,morishima17} and also pebble accretion models \citep{moriartyfischer15,johansenetal14,chambers16,johansenlambrechts17,levisonetal15pnas}.  The typical starting time of late stage of accretion simulations relates to about $\sim$3 Myr after CAI formation \citep{raymondetal09}. Most of these simulations start with fully formed giant planets and assume that the gaseous protoplanetary disk has just dissipated \cite[e.g.][]{morbidellietal12}.

The most important ingredient in terrestrial accretion models is simply the amount of available mass. A zeroth-order estimate of the Solar System's starting mass comes from the ``Minimum mass solar nebula  model'' \citep{weidenschilling77,hayashi81,desch07,crida09} (MMSN). The original MMSN model inflates the current radial mass distribution of Solar System planets to match the solar composition (adding ${\rm H}$ and ${\rm He}$; \cite{weidenschilling77,hayashi81}). Variations of this model have been proposed over the years considering the solar giant planets' orbits evolved during the Solar System history \citep{tsiganisetal05}. These models typically suggest that between the orbits of Mercury and Jupiter the primordial Solar System contained $\sim 5M_\oplus$ of solid material \citep{weidenschilling77}.

\begin{figure}
\centering
\includegraphics[scale=.65]{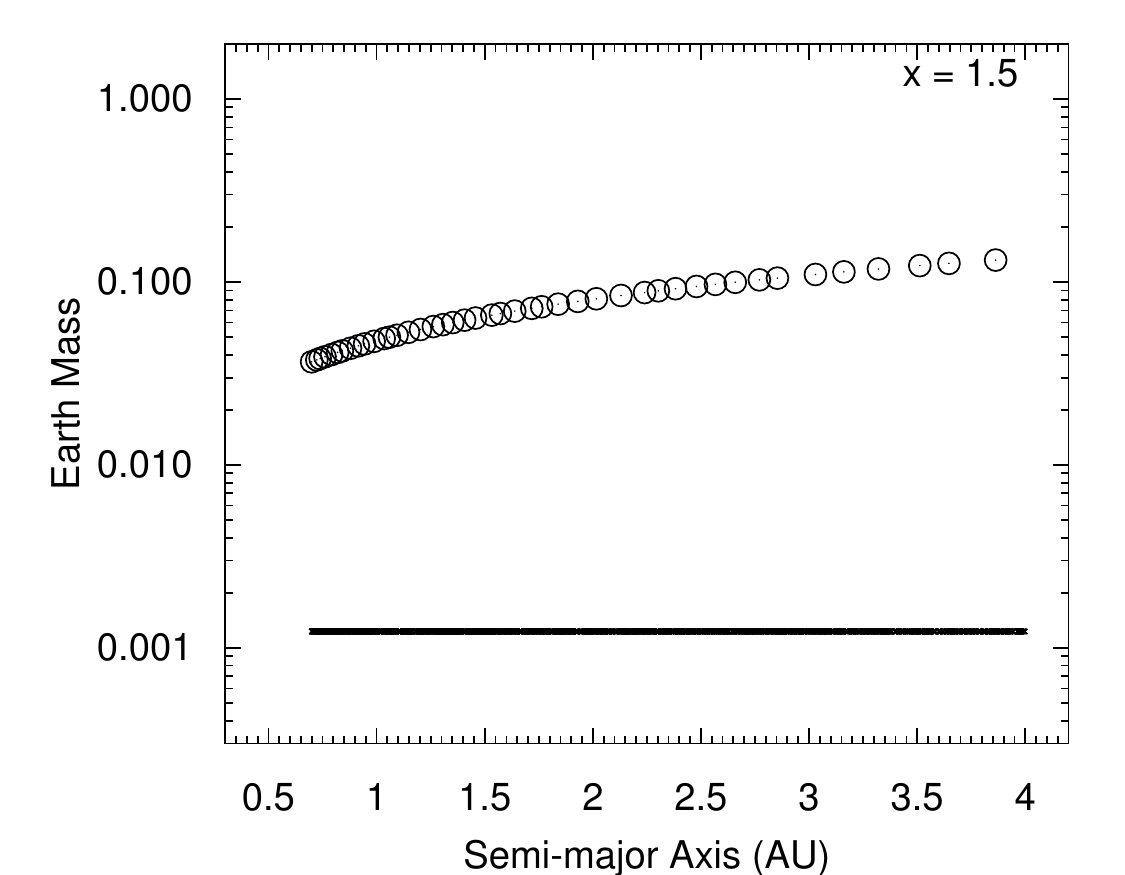}
\caption{Representative initial conditions for classical simulations of the late stage of accretion of terrestrial planets using a power-law disk. In this case x=1.5 and ${\rm \Sigma_1 =8g/cm^2}$. The mutual separation of neighbour planetary embryos is randomly selected between 5 and 10 mutual hill radii. Planetesimals are shown with masses of $\sim 10^{-3}$ Earth masses. The total mass carried by about 40 embryos and 1000 planetesimals is about 4.5${\rm M_\oplus}$.}
\label{fig:4}       
\end{figure}

Motivated by model, disk-formation simulations \citep{bate18} as well as disk observations \citep[generally of the dust component;][]{andrewsetal10,willianscieza11}, the radial distribution of solids in simulations of late stage accretion typically follow power law profiles:
\begin{equation}
{\rm \Sigma(r)=
\Sigma_1 \left(\frac{r}{1 AU}\right)^{-x}\hspace{.3cm} g/cm^2 }.
\end{equation}
${\rm \Sigma_1}$ is the surface density of solids at 1 AU. The initial mass of individual plantary embryos scales as  $r^{3(2-x)/2}\Delta^{3/2}$ \citep{kokuboida02,raymondetal05}, where x is the power-law index and $\Delta$ represents the mutual separation of adjancet planetary embryos in mutual hill radii \citep{kokuboida00}. A fraction of the disk total mass is typically distributed among planetesimals \citep{raymondetal04,raymondetal06,obrienetal06,jacobson14}. Figure \ref{fig:4} shows the distribution of planetary embryos and planetesimals in a MMSN disk profile.

At least three different scenarios for the origins of the Solar System exists. In the next Section we discuss each of them.

\subsection{The Classical Scenario}

The classical model assumes that giant planet formation can be completely separation from terrestrial planet formation. Classical model simulations simply impose a giant planet configuration (usually considering just Jupiter and Saturn) and a distribution of terrestrial building blocks. Early classical model simulations succeeded in producing a few planets in stable and well separated orbits, delivering water to Earth analogs from the outer asteroid belt and in explaining a significant degree of mass depletion of the asteroid belt~\citep{wetherill78,wetherill86,wetherill96,chamberswetherill98,agnoretal99,morbidellietal00,chambers01,raymondetal04}. Later, higher-resolution simulations were also able to match the terrestrial planets' AMD and the timing of Earth's accretion~\citep{raymondetal06,raymondetal09,obrienetal06,morishimaetal08,morishimaetal10}.

However, the classical model suffers from an important setback. Simulations systematically produce Mars analogs that are far more massive than the real planet (among other, less dramatic shortcomings). This has become known as the ``small Mars'' problem and was first pointed out by \cite{wetherill91}.  The small Mars problem is pervasive in simulations in which a) the giant planets are on low-eccentricity, low-inclination orbits, and b) the disk of terrestrial material follows a simple power-law profile with an $r^{-1}$ to $r^{-2}$ slope~\citep{obrienetal06,raymondetal06,raymondetal09,raymondetal14,morishimaetal10,izidoroetal14,izidoroetal15,fischerciesla14,kaibcowan15,haghighipourwinter16,lykawakaito17,bromleykenyon17}. 

As discussed above, a number of aspects of the Solar System can be explained if the giant planets underwent an instability known as the Nice model~\citep[originally proposed in][]{tsiganisetal05,gomesetal05,morbidellietal05}.  The original Nice model proposed that the giant planets migrated from a more compact orbital configuration to their current orbits through a late dynamical instability at about $\sim500$ Myr after CAIs formation. The instability is triggered by the gravitational interaction of the giant planets with a primordial planetesimal disk residing beyond the orbit of the giant planets. This violent dynamical event in the Solar System history is invoked to explain the late heavy bombardment, the dynamical structure of the Kuiper belt, the trojans asteroids and the potential origins of the Oort cloud.  Yet the timing of the instability is poorly-constrained.  \cite{morbidellietal18} argue that it could have happened anytime up to $\sim$500 Myr after CAIs.  

Nonetheless, only a small subset of giant planet orbits are fully self-consistent. Hydrodynamical simulations find that, during the disk phase, Jupiter and Saturn are captured in 3:2 or 2:1 mean motion resonance, usually on low-eccentricity, low-inclination orbits \citep{massetsnellgrove01,morbidellicrida07,pierensnelson08,zhangzhou10,dangelomarzari12,pierensetal14}. If the instability happened early then it is possible that Jupiter and Saturn were close to their current configuration during accretion.  If the instability happened late then the gas giants' orbits would have been much less excited during accretion.  

For a late instability the small Mars problem is insurmountable.  Mars analogs are typically as massive as Earth and embryos are often stranded in the belt~\citep{raymondetal09,fischerciesla14,izidoro15}.  There is simply no mechanism by which to deplete Mars' feeding zone.

If the instability was early, the gas giants' orbits during accretion must have been modestly more excited than they are today.  This is because their eccentricities and inclinations would have been damped below their current values from scattering of planetary embryos and planetesimals.  To end up on their correct orbits, Jupiter and Saturn must have started off with somewhat higher eccentricities ($e_J\approx e_S  \approx 0.07-0.1$). This configuration was called EEJS for `Extra Eccentric Jupiter and Saturn' by \cite{raymondetal09}. The EEJS setup is interesting because secular resonances are stronger than in the current Solar System~\citep[and much stronger than if the giant planets' orbits were near-circular][]{raymondetal09,izidoroetal16}. In classical model EEJS simulations, these secular resonances clear out Mars' feeding zone and are able to match Mars' true mass even starting with standard surface density profiles (e.g. x=1). However, it remains to be demonstrated that dynamical instabilities producing such eccentric giant planets can also satisfy other inners outer Solar System constraints and also terrestrial planets with reasonably low AMD.

\cite{hansen09} proposed that the inner Solar System had a severe mass deficit beyond 1 AU \citep{wetherill78,hansen09,morishimaetal08}. Starting from a narrow ring of embryos between 0.7 and 1. AU, Mars was scattered outside the ring beyond 1.0 AU and was starved. Hansen's simulations indeed provided a good match to the terrestrial planets. However, \cite{hansen09} did not propose a mechanism to generate a truncated disk and was not able to resolve  the asteroid belt. However, Hansen's work catalyzed the development of two scenarios for explaining the inner Solar System. The first scenario found a way to generate a truncation in the disk at 1 AU and the second further explore the implications of a primordial mass deficit beyond 1 AU. We discuss these models in the upcoming sections.

\subsection{The Grand Tack scenario}

The Grand Tack model proposes that Jupiter's migration dramatically sculpted the terrestrial planet region during the gas disk phase. The Grand Tack scenario assumes that the terrestrial planet region started with a lot of mass beyond 1 AU (~${\rm M_\oplus}$) and invokes a specific gas-driven migration history of the giant planets to deplete the region beyond 1 AU, creating a Hansen-style disk.

\begin{figure}
\hspace{-1.5cm}
\includegraphics[scale=1.5]{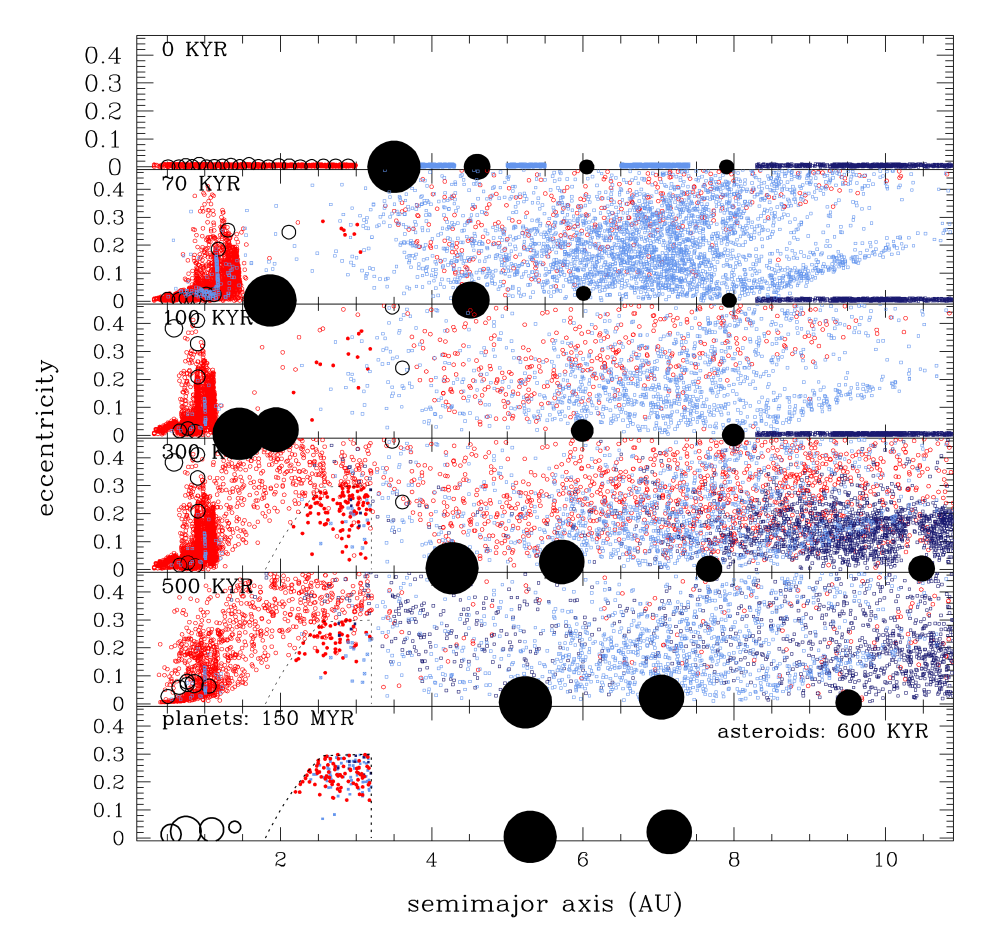}
\caption{Snapshots showing the dynamical evolution of the Solar System in the Grand Tack model. The four gas giants are represented by the black filled circles. From the innermost to the outermost one are shown: Jupiter, Saturn, Uranus and Neptune. Jupiter starts fully formed while the other giants planets grow. Terrestrial planetary embryos are represented by open circles. Water rich and water-poor planetesimals/asteroids are shown by blue and red small dots, respectively. There is a two phase of inward-then-outward migration of Jupiter and Saturn. During the inward migration phase Jupiter shepherd planetesimals and planetary embryos creating a confined disk around 1 AU. Saturn encounters Jupiter and both planets start to migrate outwards at about 100 kyr. During the outward migration phase the giant planets scatter inwards planetesimals repopulating the previously depleted belt with a mix of asteroids originated from different regions. After 150 Myr four terrestrial planets are formed. Figure from \cite{walshetal11}.} 
\label{fig:5}       
\end{figure}

The Grand Tack invokes an inward-then-outward phase of migration of Jupiter and Saturn  to sculpt the inner Solar System (see Figure \ref{fig:5}). Hydrodynamical simulations show that Jupiter is massive enough to carve a gap in the protoplanetary disk and migrate inwards in the type-II regime~\citep{lin86,ward97,durmann15}.  Saturn is less massive than Jupiter and not big enough  to carve a deep gap in the disk~\citep{crida06}. Saturn also migrates but in the very fast, type-III migration regime~\citep{massetpapaloizou03}. Migrating together in the same disk, Jupiter and Saturn typically end up in either the 3:2 or 2:1 resonance~\citep{pierensnelson08,pierensetal14}. Locked in resonance in a common gap, the balance of torques from the disk shifts and the two planets migrate outward~\citep{massetsnellgrove01,morbidellicrida07,crida09,pierenraymond11,dangelomarzari12,pierensetal14}. In the Grand Tack model, the turnaround, or ``tack point'' is set to be 1.5-2 AU as this truncates the inner disk of terrestrial material at 1 AU~\citep{walshetal11,jacobsonwalsh15,brasseretal16}.

In the Grand Tack model Jupiter and Saturn cross the asteroid belt twice. During their inward migration the giant planets shepherd most primordial asteroid material interior to 1 AU and scatter some  outwards. During their later outward migration they scatter planetesimals inward and populate the belt with a mix of planetesimals from different locations of the disk. In this model, planetesimals originally inside Jupiter's orbits are associated with the S-type asteroids (water-poor)  and planetesimals originally beyond Saturn are associated with C-Type asteroids (water-rich; e.g. \cite{walshetal12}). Some of the C-type asteroids scattered inward reach the terrestrial region and deliver water to the growing terrestrial planets~\citep{obrienetal14,obrienetal18}. After gas dispersal, the truncated narrow region around 1 AU naturally leads to the formation of good Mars analogs at 1.5 AU. Models of the subsequent Solar System evolution have indeed shown that asteroid belt produced in the Grand Tack model is consistent with the observed belt in terms of its levels of dynamical excitation and mass depletion~\citep{deiennoetal16}. The Grand Tack stands today as a viable model to explain the origins of the inner Solar System but it is not the only one.

\subsection{The primordial Low-mass/Empty asteroid belt scenario}

There are other ways that nature might produce a Hansen-style disk without invoking a dramatic giant planet migration phase. The  low-mass asteroid belt scenario proposes that the mass deficit beyond 1 AU is primordial. Perhaps solid material in the belt region drifted inside 1 AU by gas drag leaving the belt region severely mass depleted \citep{izidoroetal15,levisonetal15pnas,moriartyfischer15,drazkowska16}. This must have happened after Jupiter's core was large enough to block the inward pebble flux~\citep{lambrechts14,morby15}. The pile up scenario is also consistent with recent pebble drift and accretion models. If pebbles can drift inward from the other regions of the disk by gas drag assistance they can eventually pile up and produce disks with any surface density profile \citep{izidoroetal15}. 

In order to test which kind of disk could match the Solar System \cite{izidoroetal15} systematically studied the formation of terrestrial planets in disks with different radial mass distributions, i.e. in shallow and steep surface density profiles~\citep[][tested a much narrower range of surface density slopes]{raymondetal05,kokuboetal06}. 
 


\begin{figure}
\centering
\includegraphics[scale=.5]{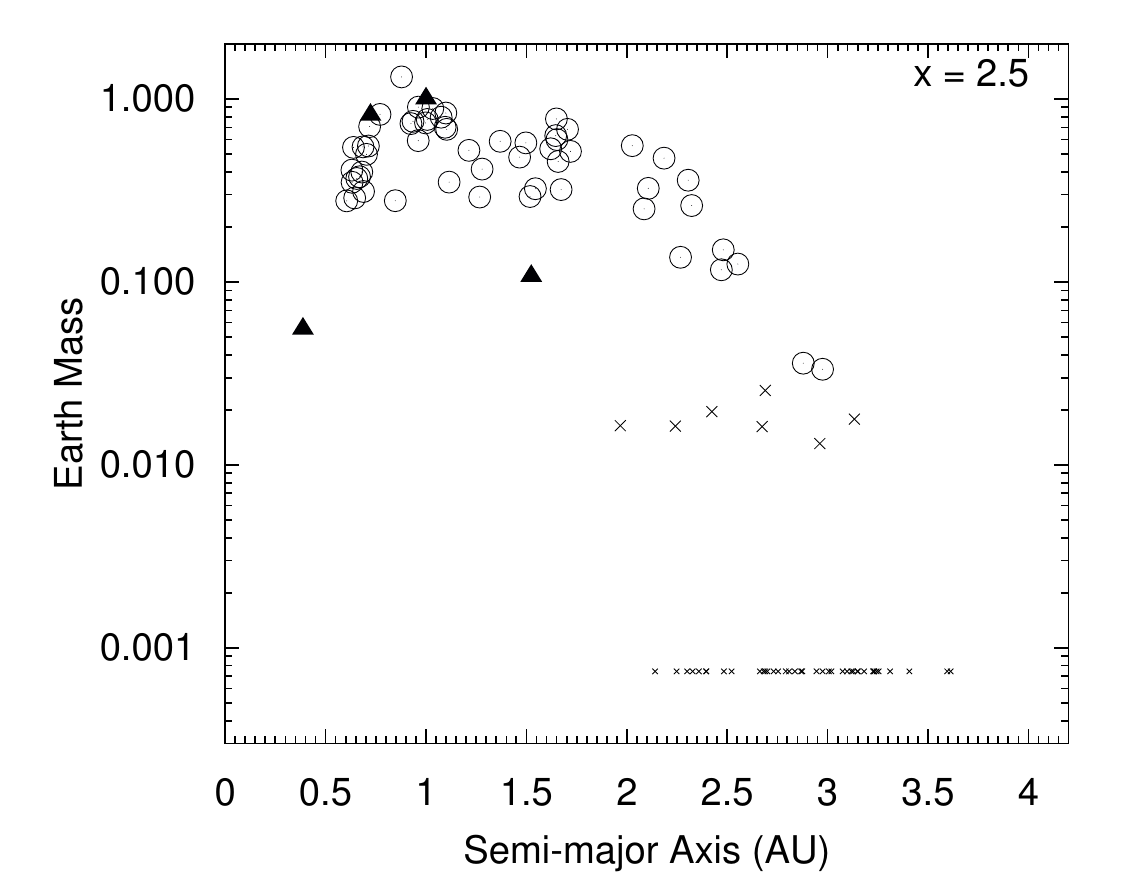}
\includegraphics[scale=.5]{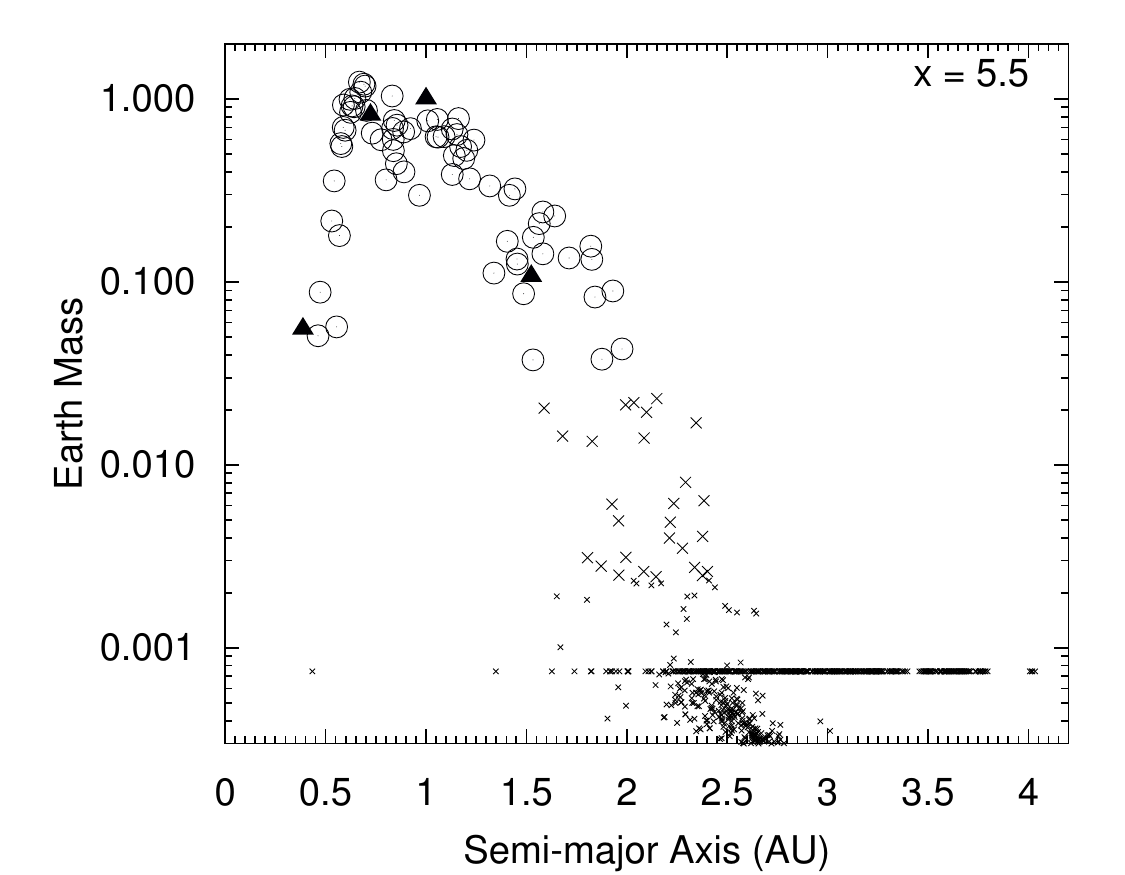}

\includegraphics[scale=.5]{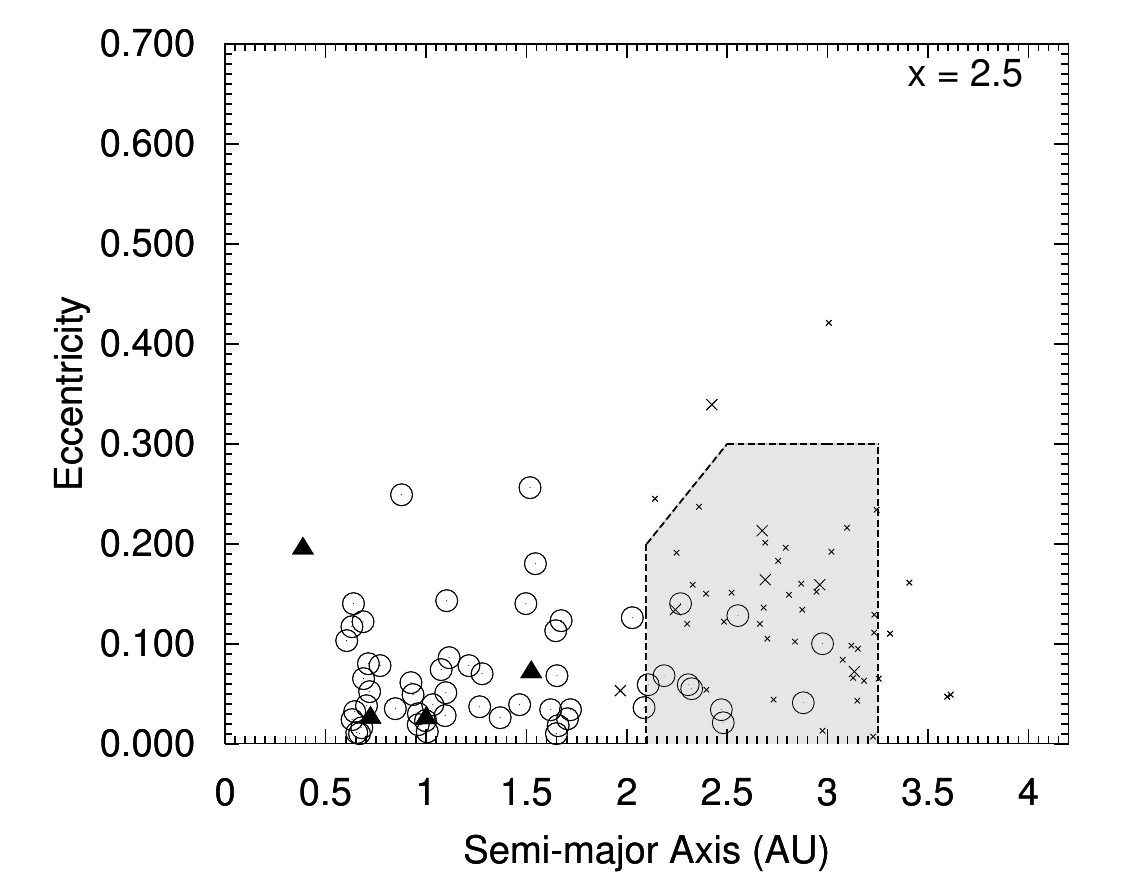}
\includegraphics[scale=.5]{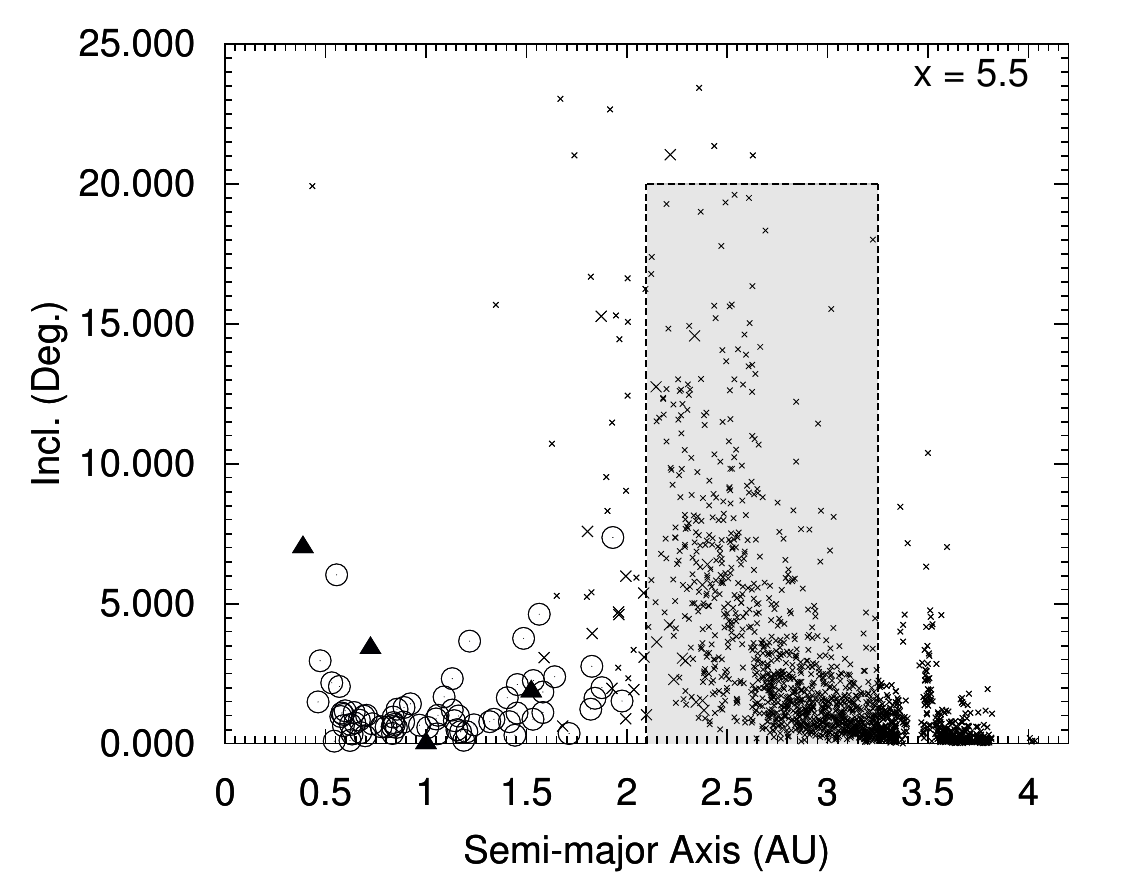}

\includegraphics[scale=.5]{asteroid_belt_e_2_5_700Myr.pdf}
\includegraphics[scale=.5]{asteroid_belt_inc_5_5_700Myr.pdf}
\caption{Final distribution of simulated planets and asteroids in simulations with x=2.5 (left-hand panels) and x=5.5 (right-hand panels) after 700 Myr of integration. The results of 15 simulations are shown for each disk. Protoplanetary objects with masses larger than 0.3 ${\rm M_{\oplus}}$ are shown with circles. Smaller bodies are labeled with crosses. The solid triangles represent the inner planets of the Solar System. Figure adapted from \cite{izidoroetal15}}
\label{fig:6}
\end{figure}

There is a trade-off between Mars' mass and the asteroid belt's level of excitation. Simulations from \cite{izidoroetal15} with shallow disks failed to produce a small Mars (see upper-left panel of Figure \ref{fig:6}). Only very steep disks with x=5.5 were successful in producing a low-mass Mars (see upper-right panel of Figure \ref{fig:6}. However, in these same simulations the asteroid belt is much dynamically colder than the real belt (see middle-right and bottom-right panels of Figure \ref{fig:6}). This level of excitation is inconsistent with the belt shown in Figure \ref{fig:3}. This is due to the severe mass deficit beyond 2 AU in steep disks, which results in a inefficient gravitational self-stirring.


To reconcile the low-mass asteroid belt scenario with the current Solar System a mechanism to excite the belt is required. \cite{izidoroetal16} showed that asteroid belt could be naturally excited to the current levels if Jupiter and Saturn's early orbits were chaotic.   If Jupiter and Saturn underwent a  phase of chaos in their orbits secular, resonances could randomly jump across the entire belt \citep{izidoroetal16} and other secular effects could be amplified (Deienno et al., in prep)  exciting asteroids across the entire belt up to the required observed levels.

\begin{figure}
\hspace{1cm}
\includegraphics[scale=.4]{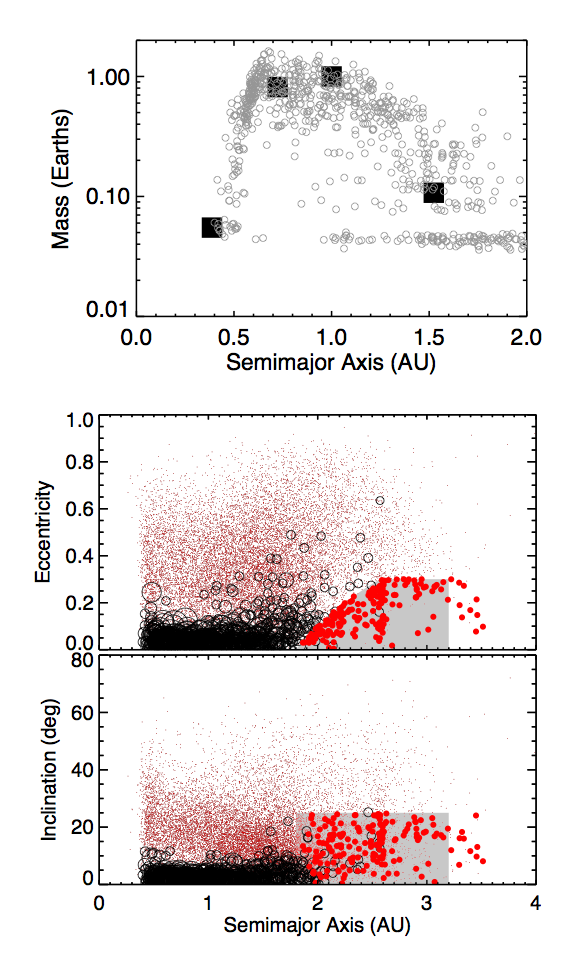}
\caption{Outcome of many simulations of the late stage of accretion of terrestrial planets in the framework of the primordial empty-asteroid belt model. The top plot shows semi-major axes versus masses of all planets formed in these simulations. The filled squares represent the real terrestrial planets. Open circles represent the simulated planets. The middle and bottom plots show semi-major axis versus eccentricity and semi-major axis vs orbital inclination, respectively. Again, planets are shown as open circles. Asteroids surviving until the end of the simulation are shown as small red dots and  S-type asteroids successfully implanted in the belt are shown by big red dots. Figure from \cite{raymondizidoro17b}} 
\label{fig:7}       
\end{figure}

\cite{raymondizidoro17b} proposed that the inner Solar System is also consistent with a primordial empty asteroid belt. Starting from a narrow annulus of embryos and planetesimals, a small fraction of planetesimals from the terrestrial region are naturally implanted into the asteroid belt. Many planetesimals are scattered onto high-eccentricity, belt-crossing orbits. A fraction of these are scattered by rogue embryos onto lower-eccentricity orbits trapped beyond 2 AU, preferentially in the inner main belt. In this model, planetesimals originated from the terrestrial are associated with S-type asteroids. Several times the current total mass in S-types is implanted in simulations that also match the terrestrial planets' masses (see Figure \ref{fig:7}) and orbits (both AMD and RMC). The empty primordial belt scenario is completely different than classical, the Grand Tack and low mass asteroid belt models in that it proposes that all S-types are refugees.

\begin{figure}
\includegraphics[scale=.5]{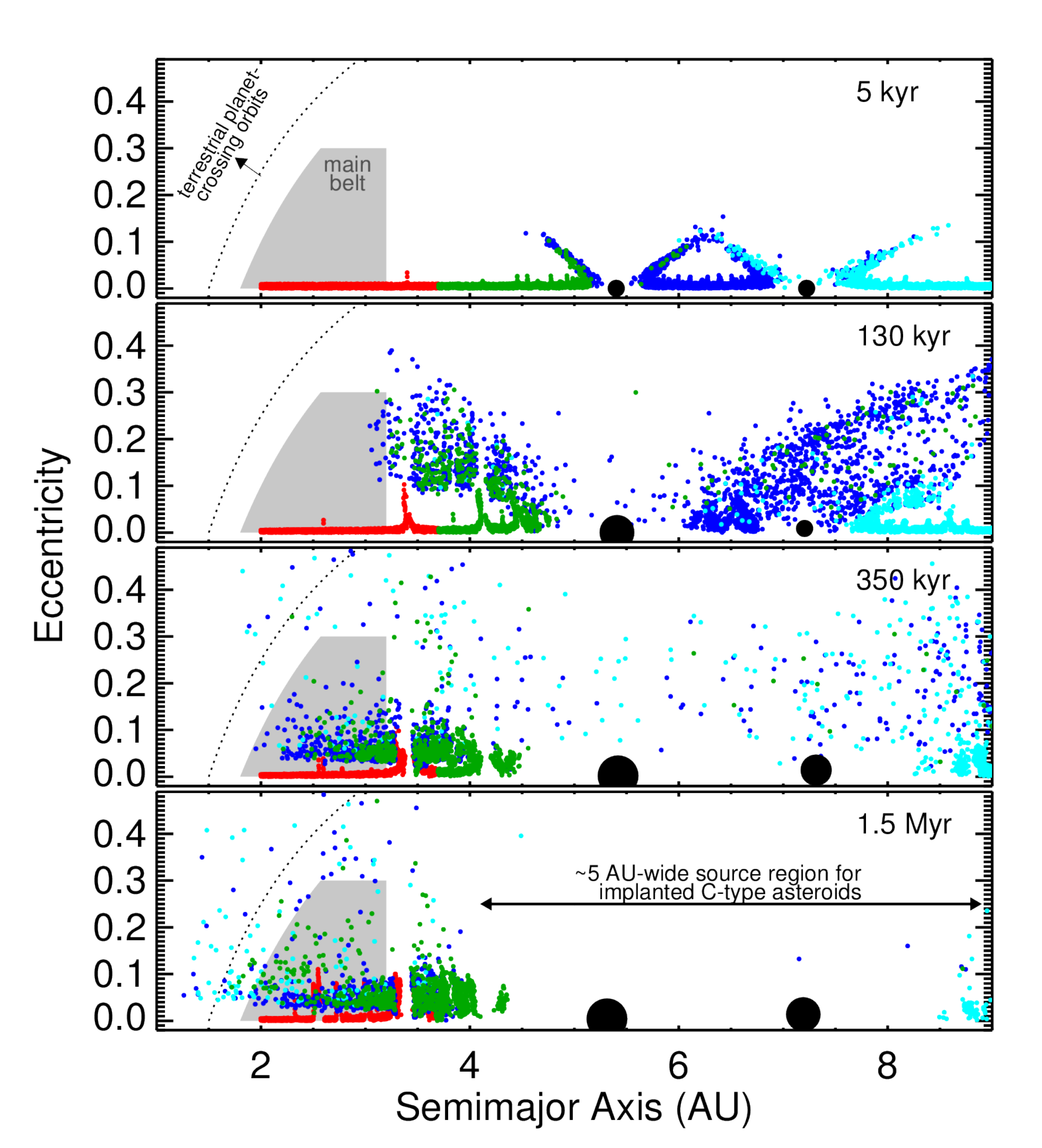}
\caption{Implantation of asteroids in the belt during the rapid gas accretion of Jupiter and Saturn. The gas giants are represented by the growing filled circles. Planetesimals are color coded to reflect their original locations.  Planetesimals have a diameter of 100~Km. The gray shaded region delimits the asteroid belt. Jupiter grew linearly from a 3${\rm M_\oplus}$ core to its current mass  from 100 to 200 kyr. Saturn start to grow later, from 300 to 400 kyr. In this simulations the planets are assumed locked in the 3:2 mean motion resonance as suggested by the results of hydrodynamical simulations. Asteroids crossing the shaded line towards the innermost parts of the inner Solar System cross the terrestrial planets orbits. Figure from \cite{raymondizidoro17a}}
\label{fig:implantation}
\end{figure}

In the primordial empty asteroid belt model the C-type asteroids are implanted in the belt by a different process. \cite{raymondizidoro17a} showed that planetesimals from  the giant planet region are inevitably implanted in the belt by gas-drag assistance during the gas disk phase. During Jupiter and Saturn's rapid gas accretion, the orbits of nearby planetesimals were perturbed and they were gravitationally scattered onto eccentric orbits. Given the dissipative nature of gas drag~\citep{adachietal76}, many planetesimals were scattered inward, had their eccentricities damped by gas drag, and were captured onto stable orbits, preferentially in the outer main belt (see Figure \ref{fig:implantation}). This mechanism is also consistent with the low-mass asteroid belt scenario.

 A fraction of asteroids is not implanted in the belt region but rather reach high eccentric orbits which crosses the terrestrial region and deliver water to the growing terrestrial planets. The formation of giant planets pollutes the inner part of the protoplanetary disk with water-rich bodies \citep{raymondizidoro17a}. In contrast with other models (e.g., the Grand Tack), this mechanism is an {\it unavoidable} consequence of giant planet formation. It must have played a role in the early Solar System.

The injection of S-types from the terrestrial planet region and C-types form the giant planet region reconcile the empty primordial asteroid belt model with the bulk of the inner Solar System architecture. Although S-type asteroids are injected in the belt in orbits which are consistent with the level of excitation of the current belt, at the end of the gas disk phase the level of dynamical excitation of asteroids implanted in the outer part of the belt (typically C-types) is far from the observed one. Thus, a mechanism to excited asteroid orbits as the chaotic excitation \citep{izidoroetal16} is still required.

 Finally, both the low-mass asteroid belt and empty asteroid belt scenarios remain as viable alternatives to the Grand-Tack model.

\subsection{Constraining and distinguishing formation scenarios}

Three current viable models of the late stage of accretion of terrestrial planet exist to explain the bulk of the inner Solar System structure. These models are consistent with the masses and orbits of the terrestrial planets (RMC and AMD), the structure of the asteroid belt and the origins of water in the inner Solar System \citep{walshetal11,izidoroetal16,raymondizidoro17a,raymondizidoro17b}. They are also self-consistent with envisioned models of the Solar System evolution  \cite{izidoroetal16,raymondizidoro17a}. While all these models seems to hold a similar degree of success obviously only one of them is potentially correct. So how can we hope to distinguish between them? 

Empirical tests to discriminate these models may be based on space observations of Solar System minor bodies \citep{morbyraymond16} or high precision isotopic measurements of different planetary objects \citep[e.g.][]{dauphas17}. From the theoretical side, models may be differentiated by more sophisticated numerical simulations studying the physical mechanisms involving pebble accretion and planetary migration \cite{izidoroetal16}. 

The Grand Tack model requires a specific large-scale giant planet migration. One of the main loose end of the Grand Tack is that it is not clear if the required inward-then-outward large scale migration is also possible in a scenario where gas accretion onto Jupiter and Saturn is self-consistently computed \citep{raymondmorbidelli14}. Unfortunately, our understanding of gas accretion onto cores is still incomplete. Hydrodynamical simulations of growth and migration of Jupiter and Saturn typically invoke a series of simplifications considering the challenge in performing high-resolution self-consistent simulations of this process \citep[e.g.][]{pierensetal14}.

The Achilles' heel of the low-mass asteroid belt model is our deficient understanding of how planetesimals form. The low-mass asteroid belt model requires planetesimals to have formed interior to 1-1.5 AU but very inefficiently in the asteroid belt, right next door. \cite{morbyraymond16} suggest that this may be achieved if planetesimals forming inside 1-1.5 AU formed much earlier than planetesimals in the belt (beyond 2 AU). Asteroids may be a low-mass second-generation of planetesimal perhaps formed during the photo-evaporation phase of the sun's gaseous disk \cite[e.g.][]{morbyraymond16}. The streaming instability requires high solid/gas ratio to operate and this could be more easily achieve during the beginning of the photo-evaporation phase \citep{carreraetal17}. This view where S-type asteroids form late is also currently supported by meteorites analysis \cite{morbyraymond16}.

It has been  suggested that at least two generations of planetesimal were born in the inner Solar System. The oldest population is associated with the parent bodies of iron meteorites, formed around half-million years after CAIs \citep{kruijeretal12}. The youngest one is associated with chondritic meteorites formed after $\sim$3~Myr \cite{villeneuveetal09}. The late formation of asteroids (potentially after the short-lived radionuclides as $^{26}$Al became an inefficient heat source) is also supported by the fact that S-type asteroids are dominated by thermally undifferentiated bodies \citep{weisselkinstanto13,scheinbergetal15}. If the terrestrial planets formed from an earlier generation of planetesimals than the S-types, this would conflict with the empty primordial asteroid belt model, in which the two populations are drawn from the same source.  It remains to be seen whether the model could still be viable in an alternate form, perhaps by invoking that the later generation of planetesimals formed at the edge of the terrestrial planet region rather than in the main belt \citep{raymondizidoro17a}.

Strong tests aiming to disentangle these models may also emerge from more complex  multidisciplinary approaches combining the accretion history of Earth produced in N-body simulations with models of core-mantle differentiation \cite{jacobsonetal14} and geochemical models \cite{dauphas17}. This may be the key to distinguish these scenarios. The amount of water delivered to the terrestrial planets in the low-mass/empty asteroid belt scenarios has not been quantified in terrestrial planet formation simulations, and confronted with those in the Grand-Tack \cite{obrienetal14}. These  subjects remain as interesting avenues for future research.

\section{Terrestrial planet formation in the context of exoplanets}

If we understand terrestrial planet formation in the Solar System (at least to some degree), then we can hopefully extrapolate to terrestrial planet formation in a more general setting.  The thousands of known exoplanets -- many of which are close to Earth-sized -- offer a testbed for our models. The difficulty is in knowing which exoplanets are truly analogous to our own terrestrial planets and which are entirely different beasts.

As of early 2018 there are more than 3,000 confirmed exoplanets. The bulk were discovered either by radial velocity surveys using Doppler spectroscopy~\citep{fischer14} or by transit surveys, notably NASA's {\it Kepler} mission~\citep{borucki10}. We now know that 10-20\% of Sun-like stars host gas giant planets~\citep{mayor11,howard12} but that hot Jupiters exist around only $\sim$1\%~\citep{wright12,howard10}. While most gas giants are found on orbits beyond 0.5-1 AU~\citep{butler06,udry07b}, the population is dominated by planets on eccentric orbits.  True Jupiter `analogs' -- with orbital radii larger 2 AU and eccentricities smaller than 0.1 -- exist but only around roughly 1\% of stars like the Sun~\citep{martin15,morbyraymond16}. This is thus an upper limit on the occurrence rate of Solar System analogs; while individual planets may be more common, similar systems to ours cannot be.

One particularly relevant, unexpected type of exoplanet are the so-called ``hot super-Earths'', often defined as being smaller than $4 \rearth$ or under $20 \mearth$ with orbits shorter than $\sim$100 days.  Super-Earths have been shown to orbit at least half of all main sequence stars, including both Sun-like~\citep{mayor11,howard12,fressin13,petigura13} and low-mass stars~\citep{bonfils13,mulders15b}. Many super-Earths are in multiple systems, which tend to have compact orbital configurations and similar-sized planets~\citep{lissauer11,lissauer11b,weiss18}. To date, up to seven have been found in the same system~\citep{gillon17,luger17}. Extensive radial velocity monitoring of {\it Kepler} super-Earths has found a dichotomy: smaller planets have high densities and are indeed rocky `super-Earths' whereas larger planets tend to have lower densities and are more likely `mini-Neptunes'~\citep{weiss13,marcy14,weiss14}.  The division between super-Earths and mini-Neptunes appears to lie close to $1.5 \rearth$~\citep{weiss14,lopez14,rogers15,wolfgang16,chen17}.  

In this section we first discuss models for the origin of super-Earths (broadly-defined to include all planets smaller than $4 \rearth$), then discuss how terrestrial planet formation may proceed in systems with gas giants on orbits very different from Jupiter's.

\subsection{Origin of super-Earth systems}

The population of super-Earths is rich enough to provide quantitative constraints on formation models:
\begin{enumerate}
\item Their occurrence rate~\citep[$\sim 50\%$ around main sequence stars;][]{mayor11,fressin13,petigura13,dong13}. 
\item Their multiplicity distribution. Systems with multiple super-Earths are much easier to confirm than single super-Earth systems because parameters must match for different planets and because transit-timing variations offer additional constraints~\citep{lissauer11b}. The observed distribution has more single super-Earth systems than multiple systems, which is sometimes referred to as the ``Kepler dichotomy''~\citep[i.e., a dichotomy between single- and multiple systems][]{fang12,tremaine12,johansen12}.
\item Their period ratio distribution, i.e., the distribution of period ratios of adjacent planets in multiple planet systems~\citep{lissauer11b,fabrycky14}. The distribution is not preferentially peaked at first order mean motion resonances neither clearly representative of a uniform distribution \citep{fabrycky14}.
\item The division between rocky super-Earths and gas-rich mini-Neptunes at $\sim 1.5 \rearth$ ~\citep{weiss14,lopez14,rogers15,wolfgang16,chen17}.
\end{enumerate}

At least 8 models have been proposed to explain the origin of super-Earths.  Before almost any super-Earths were known, \cite{raymond08} determined six potential formation pathways for super-Earths and laid out a simple framework to use observations of system architecture and planet bulk density to differentiate between them.  Several of those pathways were quickly disproven because they did not match observations; for instance, one mechanism proposed that super-Earths form from material shepherded inward by a migrating giant planet~\citep{fogg05,fogg07,raymond06b,mandell07}.  It was quickly shown that there is no correlation between close-in gas giants and super-Earths -- to the contrary, there is generally an anti-correlation between hot Jupiters and other close-in planets~\citep{latham11,steffen12}.

At the time of the writing of this chapter, two models remain viable: the {\it migration} and {\it drift} models.  Yet we think it worth clearly explaining why simple, in-situ growth of super-Earths is not a viable formation mechanism.  In-situ growth of super-Earths was first proposed by \cite{raymond08}, who subsequently discarded it based on the prohibitively large disk masses required. It was re-proposed by \cite{hansen12,hansen13} and \cite{chiang13}, and was again refuted for both dynamical and disk-related reasons~\citep{raymond14,schlichting14,schlaufman14,inamdar15,ogihara15}. The simplest argument against in-situ growth is as follows. If super-Earths form in-situ then they must grow extremely quickly because of the very dense disks required to have many Earth-masses of solids so close-in.  Yet if planets form that quickly in massive gas disks, they must migrate.  In fact, the disks required to build super-Earths close-in are so dense that aerodynamic drag acts on full-grown planets on a shorter timescale than the disk dissipation timescale~\citep{inamdar15}. Thus, in-situ growth implies that the planets must migrate.  If they migrate then their orbits change and they didn't really form ``in-situ''.  Even if super-Earths did form in-situ within a dense disk, their migration is so fast that it tends to produce a strong mass gradient in the final planet distribution, with the innermost planet always being the most massive one, which is not observed~\citep{ciardi13,weiss18}. In-situ accretion models can match observations (for example in terms of period ratio distribution) but this requires a combination of planetary systems forming in a gas free scenario and planetary systems produced in a  gas-rich environment,  where tidal eccentricity and inclination damping due to the interaction with gas drag operate but not type-I migration \citep{dawsonetal16}.  It is not clear how to decouple gas tidal damping from type-I migration for planets in the super-Earth size/mass range.

The drift model proposes that dust drifts inward but that most planet-building happens close-in.  Dust is indeed expected to coagulate and drift inward~\citep[e.g.][]{birnstiel12} and if there exists a trap very close-in, then a fraction of the mass in drifting pebbles can be captured.  \cite{chatterjee14} proposed that this trap is a pressure bump associated with the inner edge of a dead zone.  They proposed that, once a pebble ring attains a high enough density it may collapse directly into a full-sized planet.  The inner edge of the dead zone may then retreat, shifting the formation location of the next super-Earth.  This model is promising and the subject of a series of papers~\citep{chatterjee14,chatterjee15,boley14,hu16,hu17}.  However, the model is not developed to the point of being able to directly address the constraints listed above. 

\begin{figure}
\centering
\includegraphics[scale=0.5]{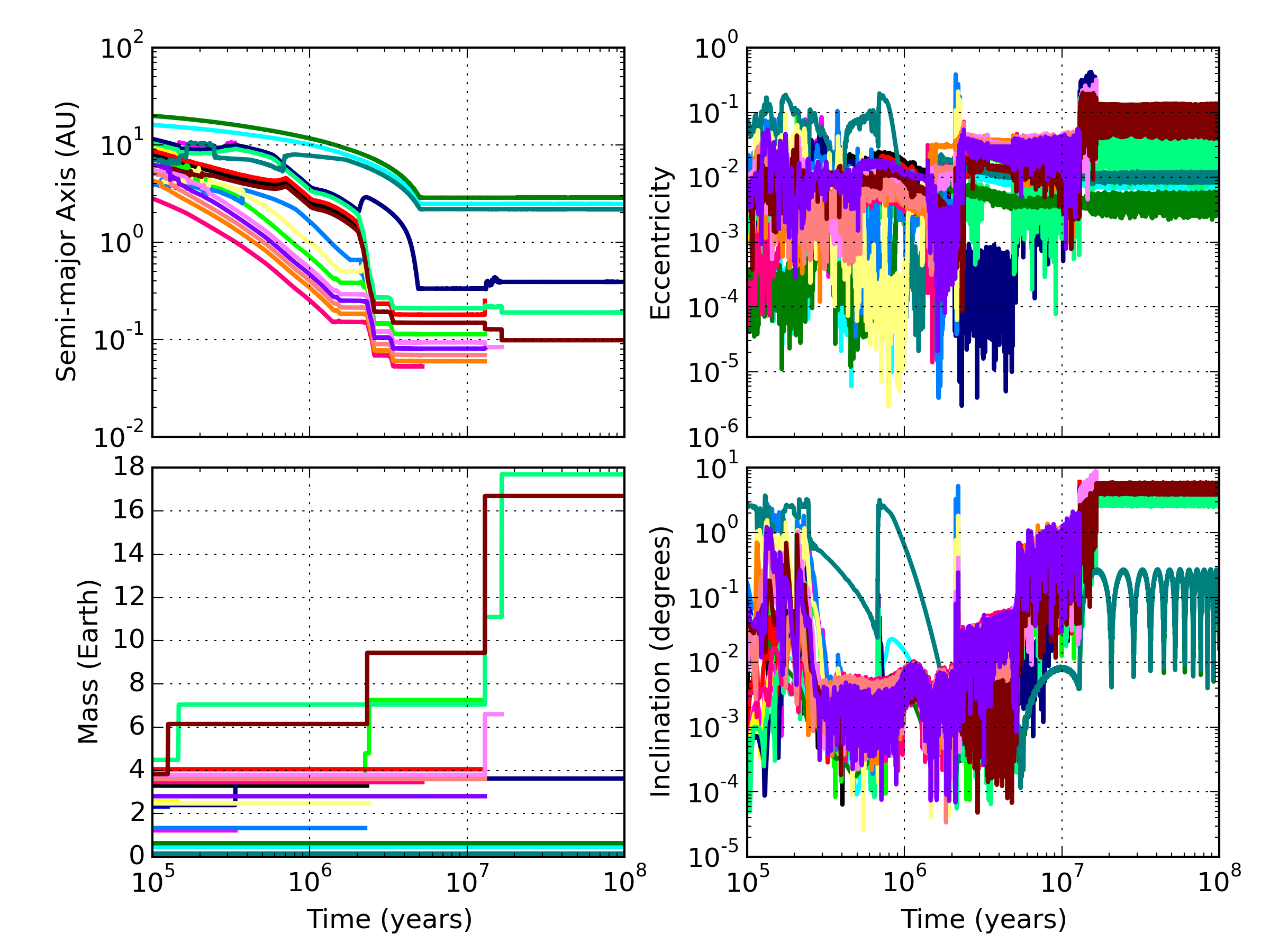}
\caption{Mass and orbital evolution of a simulation of the migration origin for systems of close-in super-Earths, from \cite{izidoro17}.  A set of $\sim$Earth-mass planetary embryos starts past the snow line and migrates inward (with occasional collisions between embryos) to create a long resonant chain, in this case consisting of 10 super-Earths interior to 0.5 AU. When the gas disk dissipates at 5 Myr, the system remained quasi-stable but underwent a large-scale instability a few Myr later (just after $10^7$ years), leading to a phase of late collisions. The final system consists of just three (relatively massive) super-Earths with modest eccentricities and large enough mutual inclinations to preclude the transit detection of all three planets. }
\label{fig:migration} 
\end{figure}

Finally, the migration model proposes that planetary embryos grow large enough far from their stars to perturb the gaseous disk and to undergo so-called type 1 migration~\citep[see example in Fig.~\ref{fig:migration};][]{goldreich80,ward86,tanaka02}. Given that disks have magnetically-truncated inner edges~\citep[e.g.][]{romanova03,romanova04}, embryos migrate inward, may be caught at planet traps \citep{lyraetal10,hasegawapudritz11,hasegawapudritz12,hornetal12,bitschetal14,alessietal17}, but eventually they reach the inner edge, where a strong torque prevents them from falling onto the star~\citep{masset06}.  Systems of migrating embryos thus pile up into chains of mean motion resonances anchored at the inner edge of the disk~\citep{cresswell07,terquem07,ogihara09,mcneil10,cossou14,izidoro14}.  Collisions are common during this phase; they destabilize the resonant chain but it is quickly re-formed. Short lived gaseous disks \citep[e.g.][]{hasegawapudritz11,bitsch15,alessietal17} or simply reduced gas accretion rates \citep{lambrechtslega17} may have frustrated the growth of sufficiently low-mass planetary embryos to gas giant planets. When the disk dissipates so too does the accompanying eccentricity and inclination damping~\citep{tanaka04,cresswell07,bitsch10}. Most resonant chains become unstable and trigger a late phase of giant collisions in a gas-free (or at least, very low gas density) environment~\citep{terquem07,ogihara09,cossou14,izidoro17}. Assuming that 5-10\% of systems remain stable after the disk dissipates, the surviving systems provide a quantitative match to both the observed super-Earth period ratio and multiplicity distributions~\citep{izidoro17}. In this context, the {\it Kepler} dichotomy is an observational artifact generated by the bimodal inclination distribution of super-Earths, a few of which have very low mutual inclinations (and thus a high probability of being discovered as multiple systems) but the majority have significant mutual inclinations generated by their late instabilities~\citep[][ see also chapter by Morbidelli]{izidoro17}.  The model simultaneously explains the existence of super-Earths in resonant chains like Kepler-223 \citep{mills16} and TRAPPIST-1~\citep{gillon17,luger17}.

How can we hope to use observations to differentiate between the drift and migration models?  In its current form the migration model is built on the assumption that embryos large enough to migrate should preferentially form far from their stars, past the snow line. In the Solar System it is indeed thought that large embryos formed in the outer Solar System and became the cores of the giant planets whereas small embryos formed in the inner Solar System and became the building blocks of the terrestrial planets~\citep{morby15}.  Given their distant formation zones, the migration model thus predicts that super-Earths should be predominantly water-rich and thus low-density~\citep{raymond08}.  In contrast, in the drift model pebbles should have time to devolatilize before accretion as they drift inward and so super-Earths should be predominantly rocky.  However, this difference depends strongly on where the first planetesimals form, as these serve as the seeds for embryo growth (in particular for pebble accretion).  This question is unresolved: some studies find that planetesimals first form at $\sim$1 AU~\citep{drazkowska16} whereas others find that planetesimals first form past the snow line~\citep{armitage16,drazkowskaalibert17,carreraetal17}.  This problem remains for further study, and it has clear implications for our interpretation of the water abundances of super-Earths.

The transition between super-Earths and mini-Neptunes is thought to be a result of a competition between accretion and erosion~\citep[][see also chapter by Schlichting]{ginzburg16,lee16}. Growing planetary embryos accrete primitive atmospheres from the disk~\citep[e.g.][]{lee14,inamdar15} but accretion is slowed by heating associated with small impacts~\citep{hubickyj05} and eroded by large impacts~\citep{inamdar16} and during the disk's dissipation~\citep{ikoma12,ginzburg16}.  Photo-evaporation of close-in planets may also play an important role, by preferentially stripping the atmospheres of low-mass, highly-irradiated planets~\citep{owen13,owen17,lopez16}. Planets located in the ``photo-evaporation valley'' -- the region of very close-in orbits where any atmospheres should have been stripped from low-mass planets -- appear to be mostly rocky~\citep{lopez17,jin18}.  Of course, this is simply for the closest-in planets, which even in the migration model may plausibly have been built from rocky material shepherded inward by migrating, volatile-rich planets~\citep{izidoro14}. Yet for more distant planets it remains challenging to determine unambiguous compositions because there are at least three categories of building blocks: rock, water and Hydrogen~\citep{selsis07,adams08}.  Only the most extreme densities can lead to a clear determination (e.g., very high density planets are likely to have little water or Hydrogen).  Many moderate-density planets can be explained either with a large water content or with a thin Hydrogen atmosphere.

The role of the central star remains to be fully incorporated into models of super-Earth formation.  Compared with FGK stars, {\it Kepler} found that M stars have more super-Earths and fewer mini-Neptunes, and for a higher total planet mass~\citep{mulders15a,mulders15b}.  This remains to be clearly understood, and may be linked with the low abundance of gas giants around M stars~\citep{lovis07,johnson07}.

\subsection{Terrestrial planet forming in systems with giant exoplanets}

We now consider how terrestrial planets may form in exoplanet systems with gas giants.  The dynamical evolution of such systems is thought to be quite different than that of Jupiter and Saturn. Indeed, the median eccentricity of giant exoplanets is 0.25~\citep{butler06,udry07b}, five times larger than that of Jupiter and Saturn.  Although observational biases preclude a clear determination, most giant exoplanets are located somewhat closer to their stars, typically at 1-2 AU~\citep{cumming08,mayor11,rowan16,wittenmyer16}.

Two key processes are thought to be responsible for shaping the orbital distribution of giant exoplanets: inward (type 2) migration and planet-planet scattering. While Jupiter and Saturn certainly migrated, the extent of migration remains unclear.  Among exoplanet systems a much wider variety of outcomes is possible, as some planets may have migrated all the way to the inner edge of the disk to become hot Jupiters~\citep{lin86,lin96}.  The migration timescale depends on the disk's properties and is typically hundreds of thousands to millions of years~\citep{ward97,papaloizou06,kley12,durmann15} The high eccentricities of giant exoplanets are easily explained if the observed planets are the survivors of system-wide instabilities during which giant planets scattered repeatedly off of each other during close but violent passages inside each others' Hill spheres~\citep{rasio96,weidenschilling96,lin97,adams03,moorhead05,ford03,chatterjee08,ford08,raymond08,raymond10}.  This phase of planet-planet scattering typically concludes with the ejection of one of more planets. In some cases scattering can push planets to such high eccentricities that they pass very close to their stars at pericenter, and tidal dissipation can circularize and shrink their orbits, thus providing an alternate channel for the origin of hot Jupiters~\citep{nagasawa08,beauge12}.

In light of our current understanding, we now ask the question: how do giant planet migration and scattering affect the growth and evolution of terrestrial planet formation?

Giant planet migration has been shown to be much less destructive to terrestrial planet formation than was generally assumed in the late 1990s and early 2000s~\citep{gonzalez01,lineweaver04}.  An inward-migrating gas giant does not simply collide with the material in its path~\citep[except in rare circumstances;][]{tanaka99}.  Rather, strong inner mean motion resonances acting in concert with gas drag shepherd material inward, catalyzing the formation of planets interior to the giant planets' final orbits~\citep{fogg05,fogg07,raymond06b,mandell07}.  A significant amount -- typically $\sim$50\% for typical disk parameters -- of material undergoes close encounters with the giant planet during its migration and is scattered outward and stranded on eccentric and inclined orbits as the giant planet migrates away.  This material re-accumulates into a new generation of terrestrial planets that tend to have extremely wide feeding zones and are thus very volatile-rich~\citep{raymond06b,mandell07}.

The eccentricity distribution of eccentric giant planets can be matched by  planet-planet scattering models \citep[e.g.][]{chatterjeeetal08,jurictremaine08,fordrasio08,raymondetal08}. Extrasolar giants planets are typically very eccentric with median eccentricity of about  0.25 \citep{butler06,udry07b}. About 90\% of the gas giants inside 1 AU have eccentricities larger than 0.1~.
In contrast to giant  planet migration, giant planet scattering is typically very destructive to terrestrial planet formation.  When gas giants go unstable they scatter each other onto eccentric orbits, and any small bodies (planetesimals, planetary embryos or planets) in their path are typically destroyed~\citep{veras05,veras06,raymond11,raymond12,matsumura13,marzari14,carrera16}.  Objects that are closer-in than the gas giants are preferentially driven onto such eccentric orbits that they collide with the host star, whereas more distant objects are more likely to be ejected~\citep{raymond11,raymond18,marzari14}. There is an interesting observational consequence of this process.  Both outer planetesimal belts and the terrestrial planet region are strongly perturbed by the giant planets in between.  Debris disk may results from ongoing collisions of planetesimals in these outer regions ~\citep[][see also review chapter by Wyatt]{wyatt08,krivov10}. This produces a theoretically-anticipated correlation between debris disks  and low-mass planets~\citep{raymond11,raymond12}.  Such a correlation has not yet been detected~\citep{moromartin15} but surveys are ongoing.

Given that eccentric orbits are the norm among giant exoplanets, this raises the question: Why are not Jupiter and Saturn also very eccentric ones? As discussed above, our current view of Solar System evolution invokes a dynamical instability (planet-planet scattering) among the giant planets in the Solar System history \citep{gomesetal05,levisonetal11,nesvornymorbidelli12,deiennoetal17}. However, simulations that match the Solar System show a clear trend: they avoid close encounters between Jupiter and Saturn during the instability phase \citep{morbidellietal07}. Close encounters between a gas- and ice-giant or between two ice giants are common, but not between two gas giants \citep{morbidellietal07}. When a Jupiter-Saturn encounter does happen, Saturn is typically ejected from the system and Jupiter survives on a much more eccentric orbit than its present-day one. This less dramatic instability may have prevented our terrestrial planets from being destroyed, although even a weaker instability can also put the terrestrial planets at risk \citep{brasseretal09,kaibchambers16}.


\section{Putting our Solar System in context}

We now ask some big questions. How can we understand our Solar System in a larger context?  What are the key processes that make our system different than most?  Did our Solar System once host a system of hot super-Earths? (quick answer to the last question: no).

Jupiter is likely the Solar System's primary architect. Let us consider its potential effects on the growth of other planets at different phases of growth. Jupiter's core was perhaps seeded by an early generation of planetesimals that then grew by pebble accretion~\citep[][but see also Brouwers et al. (2017) and Alibert (2017) ]{ormeletal10,lambrechts12,lambrechts14}.  It is unclear {\it where} this took place. Studies have covered the full spectrum of possibilities, from distant formation followed by inward migration~\citep{bitsch15} to in-situ growth~\citep{levisonetal15nat} to close-in formation followed by outward migration~\citep{raymond16}.  Nonetheless, once its core reached $\sim 20 \mearth$ it created a pressure bump exterior to its orbit that blocked the inward pebble flux~\citep{lambrechtsetal14}. This acted to starve the inner Solar System and may contribute to explaining why the embryos in the inner Solar System were so much smaller than the large cores in the Jupiter-Saturn region~\citep{morby15}. Although the direction and speed are uncertain, Jupiter's core subsequently migrated, shepherding any nearby cores and planetesimals~\citep{izidoro14}. When Jupiter underwent rapid gas accretion it strongly perturbed the orbits nearby small bodies, scattering them across the Solar System~\citep[and implanting some in the inner Solar System][]{raymondizidoro17a}.  It carved a gap in the disk and transitioned to slower, type 2 migration~\citep{lin86,ward97,crida06}.  Jupiter now provided a strong barrier for more distant planetary embryos that would otherwise migrate inward to become close-in super-Earths~\citep{izidoro15}.  Blocked by Jupiter and Saturn, these embryos instead accreted to form the ice giants~\citep{izidoro15b}.  Once the disk dissipated, Jupiter's dynamical influence played a key role in the late-stage accretion of the terrestrial planets and the dynamical sculpting of the asteroid belt~\citep[e.g.][]{raymondetal14,morbyraymond16}.

There are thus two potential ways that Jupiter may explain why the Solar System is different, specifically our lack of super-Earths.  The first is by blocking the pebble flux and starving the growing terrestrial planetary embryos. The second is by blocking the inward migration of large cores~\citep{izidoro15b}.  However, it is worth noting that two studies have proposed that the Solar System once contained a population of super-Earths that was later destroyed~\citep{volk15,batygin15}.  Let us consider whether this is plausible.

\begin{figure}
\hspace{-1.3cm}
\includegraphics[scale=.5]{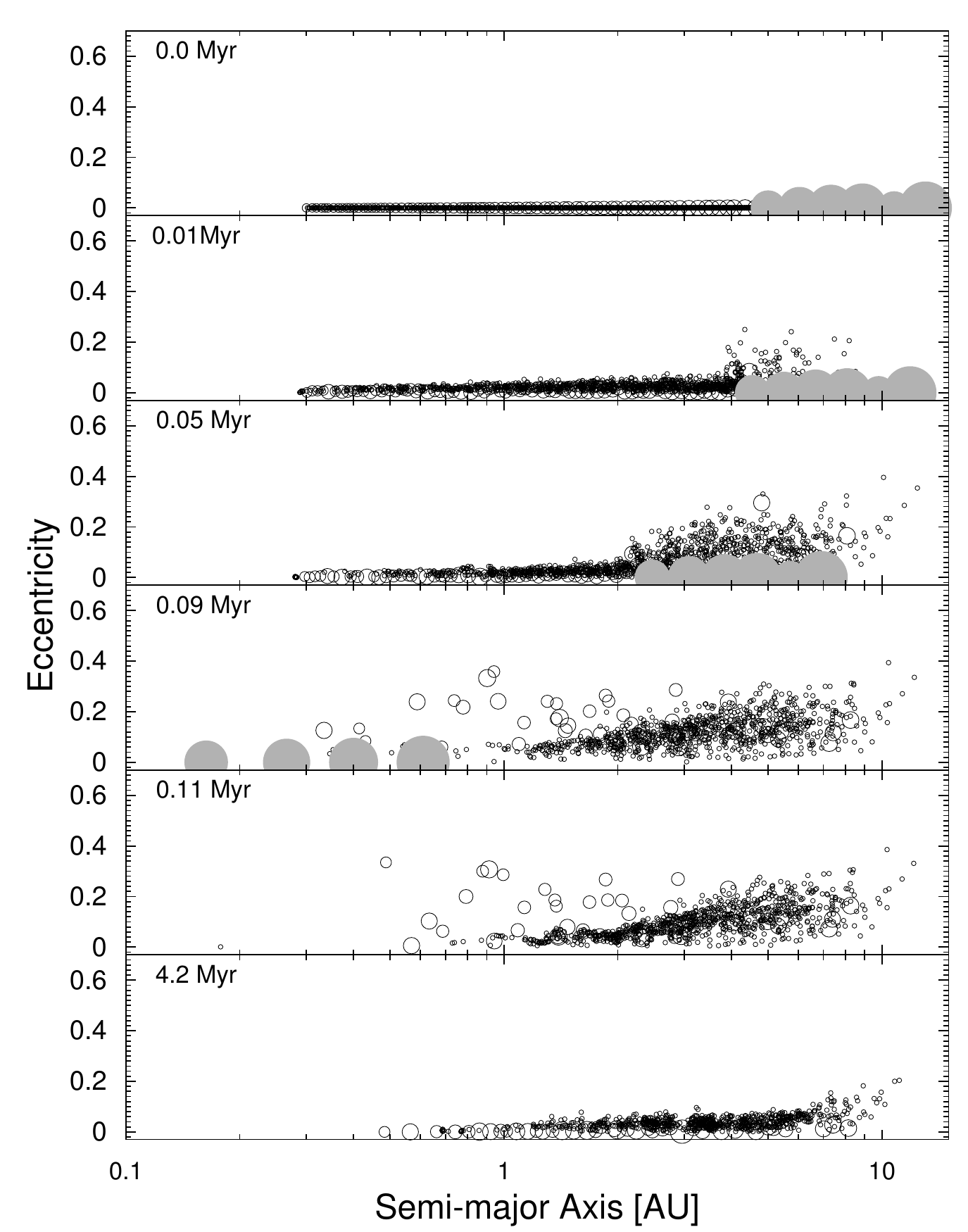}
\includegraphics[scale=.5]{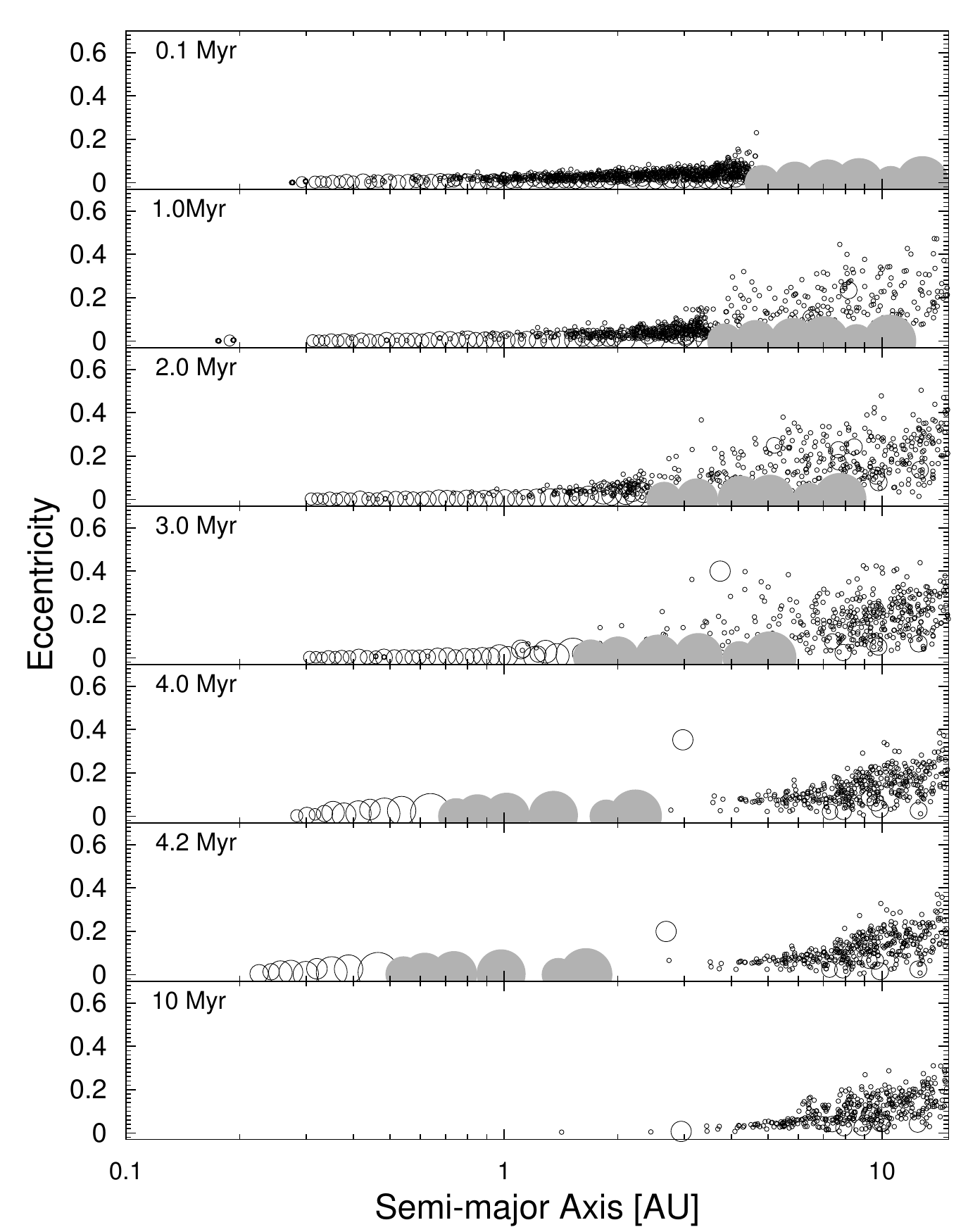}
\caption{Snapshots  of the dynamical evolution of a population of planetesimal and planetary embryos in the presence of migrating super-Earths. The gray filled circles represent the super-Earths. planetary embryos and planetesimals are shown by open circles and small dots, respectively. The super-Earth system is composed of six super-Earths  with masses roughly similar to those of the Kepler 11 system \citep[e.g.][]{lissaueretal13}. The left-panel shows a simulation where the system of super-Earths migrate fast, in a short timescale of about 100~kyr. The right-panel represents a simulation where the system of super-Earths migrate slowly, in a timescale comparable to the disk lifetime. Figure adapted from \citep{izidoro14}}
\label{fig:migratingSE}
\end{figure}

If the Solar System's presumed primordial super-Earths formed by migration, they must have migrated inward {\it through} the building blocks of the terrestrial planets. Type-I migration may be directed inwards or outwards (see chapter by Nelson) or even be halted depending on the disk local properties \citep{ward86,ward97a,paardekoopermellema06,paardekoopermellema08,baruteaumasset08,paardekooperpapaloizou08,kleyetal09,kleycrida08,paardekooperetal10,paardekooperetal11}. However, as the disk evolves and cools down any type-I migrating planets are eventually released to migrate inwards \citep{lyraetal10,hornetal12,bitschetal14}. If their migration was slow, the super-Earths would have swept the region around 1 AU clean of rocky material such that any planets that formed there would be decidedly un-Earth-like~\citep[see right-panel of Figure \ref{fig:migratingSE}][]{izidoro14}.  However, if their migration was fast, super-Earths would simply migrate past rocky planetary embryos without completely disrupting their distribution (see left-panel of Figure \ref{fig:migratingSE}). Alternately, if super-Earths formed by the drift model, it is plausible that they could have accumulated material close-in without perturbing the terrestrial planets' growth.  Thus, the growth of a population of close-in super-Earths in the Solar System seems plausible.

The next question is: what could have happened to such a population of close-in super-Earths?  \cite{volk15} proposed that they were ground to dust by a series of giant erosive collisions.  \cite{batygin15} proposed instead that Jupiter's migration led to a spike of collisional grinding at $\sim$1AU that produced a population of small ($\sim$100-m) planetesimals that drifted inward and became trapped in exterior resonances with the super-Earths.  Strong aerodynamic dissipation in the planetesimals' orbits pushed the super-Earths onto the young Sun.

While provocative, each of these studies neglects the fact that planet-forming disks have inner edges~\citep[e.g., see magneto-hydrodynamic simulations of][]{romanova03,romanova08}. These edges prevent planets or debris from simply falling onto the Sun~\citep[see discussion in][]{raymond16}. Rather, all studies to date suggest that processes that generate dust or debris should rather catalyze the further growth of close-in planets~\citep{leinhardt09,kenyon09,chatterjee14}.  If super-Earths indeed formed in the Solar System, they should still be here. We conclude that there is no compelling evidence that super-Earths ever formed in the Solar System.

Considering only the Sun and Jupiter, exoplanet statistics tell us that the Solar System is already at best a 1\% outlier~\citep[and more like 0.1\% when considering all stellar types; see discussion in][]{morbyraymond16}.  Yet it is likely that Earth-sized planets on Earth-like orbits may be far more common.  The Drake equation parameter eta-Earth -- the fraction of stars that host a roughly Earth-mass or Earth-sized planet in the habitable zone -- has been directly measured for low-mass stars to be tens of percent~\citep{bonfils13,kopparapu13,dressing15}.  Yet how `Earth-like' are such planets?  Without Jupiter, would a planet at Earth's distance still look like our own Earth? 

When viewed through the lens of planet formation, two of Earth's characteristics are unusual: its water content and formation timescale.  The building blocks of planets tend to either be very dry or very wet ($\sim$10\% water like carbonaceous chondrites, or $\sim$50\% water like comets). While Earth's composition can be explained by having grown mostly from dry material with only a sprinkling of wet material.  A simple explanation is that, even though its formation provided a sprinkling of water-rich material~\citep{raymondizidoro17a}, Jupiter blocked later water delivery \cite[e.g.][]{morbidellietal16,satoetal16}.  Without Jupiter it stands to reason that Earth should either be completely dry or, more likely, much wetter. 

Earth's last giant impact is constrained not to have happened earlier than $\sim$40 Myr after CAIs~\citep{toubouletal07,kleineetal09,aviceetal17}. However, most `Earth-like' planets probably form much faster.  Super-Earths typically complete their formation shortly after dispersal of the gaseous disk~\citep{izidoro17,alessietal17}. Accretion in the terrestrial planet zone of low-mass stars is similarly fast whether or not migration is accounted for~\citep{raymond07,lissauer07,ogihara09}. The geophysical consequences of fast accretion remain to be further explored, but it stands to reason that fast-growing planets are likely to be hotter and may thus lose more of their water compared with slower-growing planets like Earth.  This could in principle counteract our previous assertion that most terrestrial planets should be wetter than Earth.

While other Earths remain a glamorous target for exoplanet searches, we think that understanding how other planets are similar to and different than our own Solar System is a worthy goal in itself.  For instance, the abundance and configuration of ice giants on orbits exterior to gas giants will constrain our understanding of orbital migration.  Likewise, the radial ordering of systems with different-sized planets at different orbital distances will constrain models of pebble accretion.

\section{Summary}

We have reviewed the current paradigm of terrestrial planet formation, from dust-coagulation to planetesimal formation to the late stage accretion. We discussed the classical scenario of terrestrial planet formation and the more recently proposed alternatives to the Grand-Tack model, the primordial low-mass and empty asteroid belt models. We discussed the origins of hot super-Earths, placed the Solar System in the context of exoplanets and discussed terrestrial planet formation in exoplanetary systems. Below we summarize some of the key questions discussed here:
\begin{itemize}

\item The streaming instability stands as a promising mechanism to explain how mm- to cm-sized particles grow to 100 km-scale planetesimals. Yet the streaming instability require specific conditions to operate and this implies that planetesimals may form in preferential locations (e.g., just beyond the snow line).

\item Planetesimals grow into planetary embryos (or giant planet cores) by accreting planetesimals or pebbles (or a combination of both). Simulations of planetesimal accretion struggle to grow giant planet cores within the lifetime of protoplanetary disks. Pebble accretion may solve this long-standing timescale conflict, but key aspects of pebble accretion remain to be better understood.

\item Three models of the late stage of accretion of terrestrial planet can explain the structure of the inner Solar System: the Grand-Tack, the primordial low-mass asteroid belt and the primordial empty asteroid belt scenarios. A clear future step in planet formation is to differentiate between these models. Tests may be based on observations or detailed studies of key mechanisms such as the location of planetesimal formation, gas accretion onto cores and planet migration. Combining N-body simulations with geochemical models is another powerful tool.

\item  Hot super-Earths cannot form by pure in-situ accretion. Super-Earths forming in-situ would grow extremely fast because of the large solid masses required in the inner regions and the corresponding short dynamical timescales. If super-Earths form rapidly in the gaseous disk, they must migrate and not form `in-situ'.

\item Close-in super-Earths may have formed farther from their stars and migrated inward. Migration creates resonant chains anchored at the inner edge of the disk, most of which destabilize when the disk dissipates and quantitatively match the super-Earths' observed properties. No system of super-Earths is likely to have formed in the Solar System simply because it should still exist today (given that disks have inner edges that prevent planets from migrating onto their stars).  

\item The Solar System is quantifiably unusual in its lack of super-Earths and in having a wide-orbit gas giant on a low-eccentricity orbit (a $\sim$1\% rarity among Sun-like stars).  These two characteristics may be linked, as Jupiter may have prevented Uranus and Neptune from invading the inner Solar System. In addition, the lack of close encounters between Jupiter and Saturn during the Solar System instability may have prevented the destruction of the terrestrial planets. 

\item Future exoplanet surveys proving data on the occurrence of planets at moderate distances from the host star and more refined constraints on the bulk composition of transiting low-mass planets will shed light on the deep mysteries of terrestrial planet formation.

\end{itemize}

\begin{acknowledgement}
We acknowledge a large community of colleagues whose contributions made this review possible. A. I. thanks FAPESP (S\~ao Paulo Research Foundation) for support via grants 16/12686-2 and 16/19556-7. S. N. R. thanks the Agence Nationale pour la Recherche via grant ANR-13-BS05-0003-002 (MOJO). We thank Ralph Pudritz for the invitation to write this review. A. I. is also truly grateful to doctor Marcelo M. Sad for his dedication, calm and expertise during the treatment of a health problem manifested during the preparation of this project.

\end{acknowledgement}

\bibliographystyle{spbasicHBexo}  
\bibliography{HBexoTemplateBib} 

\end{document}